\documentstyle[prl,multicol,aps,epsfig]{revtex}

\begin{document}

\begin{title}
{\bf Discrete gap solitons in a diffraction-managed
waveguide array}
\end{title}

\author{P.G. Kevrekidis$^1$, B.A. Malomed$^2$ and Z. Musslimani$^3$}
\address{$^{1}$Department of Mathematics and Statistics, 
University of Massachusetts, Amherst MA 01003-4515, USA \\
$^{2}$Department of Interdisciplinary Studies, Faculty of
Engineering, Tel Aviv University, Tel Aviv, Israel \\
$^{3}$Department of Applied Mathematics, University of Colorado,
Campus Box 526, Boulder, CO 80309-0526, USA\\
}
\date{\today}
\maketitle




\begin{abstract}

A model including two nonlinear chains with linear and nonlinear couplings
between them, and opposite signs of the discrete diffraction inside the
chains, is introduced. In the case of the cubic [$\chi ^{(3)}$]
nonlinearity, the model finds two different interpretations in terms of
optical waveguide arrays, based on the diffraction-management concept. A
continuum limit of the model is tantamount to a dual-core nonlinear optical
fiber with opposite signs of dispersions in the two cores. Simultaneously,
the system is equivalent to a formal discretization of the standard model of
nonlinear optical fibers equipped with the Bragg grating. A straightforward
discrete second-harmonic-generation [$\chi ^{(2)}$] model, with opposite
signs of the diffraction at the fundamental and second harmonics, is
introduced too. Starting from the anti-continuum (AC) limit, soliton
solutions in the $\chi ^{(3)}$ model are found, both above the phonon band
and inside the gap. Solitons above the gap may be stable as long as they
exist, but in the transition to the continuum limit they inevitably
disappear. On the contrary, solitons inside the gap persist all the way up
to the continuum limit. In the zero-mismatch case, they lose their stability
long before reaching the continuum limit, but finite mismatch can have a
stabilizing effect on them. A special procedure is developed to find
discrete counterparts of the Bragg-grating gap solitons. It is concluded
that they exist all the values of the coupling constant, but are stable only
in the AC and continuum limits. Solitons are also found in the $\chi ^{(2)}$
model. They start as stable solutions, but then lose their stability. Direct
numerical simulations in the cases of instability reveal a variety of
scenarios, including spontaneous transformation of the solitons into
breather-like states, destruction of one of the components (in favor of the
other), and symmetry-breaking effects. Quasi-periodic, as well as more
complex, time dependences of the soliton amplitudes are also observed as a
result of the instability development.

\end{abstract}

\begin{multicols}{2}

\section{Introduction}

\subsection{Objectives of the work}

Solitary-wave excitations in discrete nonlinear dynamical models (lattices)
is a subject of great current interest, which was strongly bolstered by
experimental observation of solitons in arrays of linearly coupled optical
waveguides \cite{Yaron1} and development of the diffraction management (DM)
technique, which makes it possible to effectively control the discrete
diffraction in the array, including a possibility to reverse its sign (make
the diffraction anomalous) \cite{Yaron2,Falk}. It has recently been shown
that a lattice subject to periodically modulated DM can also support stable
solitons, both single-component ones \cite{Ziad1,Peschel} and two-component
solitons with nonlinear coupling between the components via the
cross-phase-modulation (XPM) \cite{Ziad2}.

Two-component nonlinear-wave systems, both continuum and discrete, which
feature a {\em linear} coupling between the components, constitute a class
of media which can support gap solitons (GSs). A commonly known example of a
continuum medium that gives rise to GSs is a nonlinear optical fiber
carrying a Bragg grating \cite{gap1,gapsolitons}, whose standard model is
based on the equations 
\begin{equation}
\begin{array}{l}
i\Psi _{t}+i\Psi _{x}+\left( |\Psi |^{2}+2|\Phi |^{2}\right) \Psi +\Phi =0,
\\ 
i\Phi _{t}-i\Phi _{x}+\left( |\Phi |^{2}+2|\Psi |^{2}\right) \Phi +\Psi =0,
\end{array}
\label{BG}
\end{equation}
where $\Psi (x,t)$ and $\Phi (x,t)$ are amplitudes of the right- and
left-propagating waves, and the Bragg-reflection coefficient is normalized
to be $1$. Another optical system that may give rise to GSs is a dual-core
optical fiber with asymmetric cores, in which the dispersion coefficients
have opposite signs \cite{Dave}.

In this work, we demonstrate that the use of the DM technique provides for
an opportunity to build a double lattice in which two discrete subsystems
with {\em opposite} signs of the effective diffraction are linearly coupled,
thus opening a way to theoretical and experimental study of discrete GSs, as
well as of solitons of different types (solitons in linearly coupled
lattices with identical discrete diffraction in the two subsystems have
recently been considered in Ref. \cite{Demetri}; a possibility of the
existence of discrete GSs in a model of a nonlinear-waveguide array
consisting of alternating cores with two different values of the propagation
constant was also considered recently \cite{SK}). The objective of the work
is to introduce this class of systems and find fundamental solitons in them,
including the investigation of their stability. We will also consider, in a
more concise form, another physically relevant possibility, viz., a discrete
system with a second-harmonic-generating (SHG) nonlinearity, in which the
diffraction has opposite signs at the fundamental and second harmonics.
Solitons will be found and investigated in the latter system too.

It is relevant to start with equations on which our $\chi ^{(3)}$ model (the
one with the cubic nonlinearity) is based, 
\begin{eqnarray}
i\frac{d\psi _{n}}{dt} &=&-\left( C\Delta_2 +q\right) \psi _{n}-\left( \left|
\psi _{n}\right| ^{2}+\beta \left| \phi _{n}\right| ^{2}\right) \psi
_{n}-\kappa \phi _{n}=0,  \label{zeq1} \\
i\frac{d\phi _{n}}{dt} &=&\delta \cdot \left( C\Delta_2 +q\right) \phi
_{n}-\left( \left| \phi _{n}\right| ^{2}+\beta \left| \psi _{n}\right|
^{2}\right) \phi _{n}-\kappa \psi _{n}=0,  \label{zeq2}
\end{eqnarray}
where $\psi _{n}(t)$ and $\phi _{n}(t)$ are complex dynamical variables in
the two arrays (sublattices), $\kappa $ and $\beta $ being coefficients of
the linear and XPM coupling between them, and $t$ is actually not time, but
the propagation distance along waveguides, in the case of the most
physically relevant optical interpretation of the model. The operators 
$C\Delta_2 \psi _{n}\equiv C\left( \psi _{n+1}+\psi _{n-1}-2\psi _{n}\right)$
and $\left( -\delta \right) C\Delta _{2}\phi _{n}\equiv \left( -\delta
\right) C\left( \phi _{n+1}+\phi _{n-1}-2\phi _{n}\right) $ represent
discrete diffraction induced by the linear coupling between waveguides
inside each array, the diffraction being normal in the first sublattice and
anomalous in the second, with a {\em negative} relative diffraction
coefficient $-\delta $ and intersite coupling constant $C$ (one may always
set $C>0$, which we assume below). Physical reasons for having $-\delta <0$
are explained below. Finally, the real coefficient $q$ accounts for a
wavenumber mismatch between the sublattices.

We also choose a similar SHG model, following the well-known pattern of
discrete SHG systems with normal diffraction at both harmonics \cite
{discrSHG}, \cite{SHGreview}: 
\begin{eqnarray}
i\frac{d\psi _{n}}{dt} &=&-C\Delta _{2}\psi _{n}-\psi _{n}^{\star }\phi _{n},
\label{jeq12} \\
2i\frac{d\phi _{n}}{dt} &=&\delta C\Delta _{2}\phi _{n}-\psi _{n}^{2}-\kappa
\phi _{n},  \label{jeq13}
\end{eqnarray}
where the asterisk stands for the complex conjugation and $\kappa $ is a
real mismatch parameter. In this case too, we assume $-\delta <0$.

\begin{figure}[tbp]
\epsfxsize=7.5cm
\centerline{\epsffile{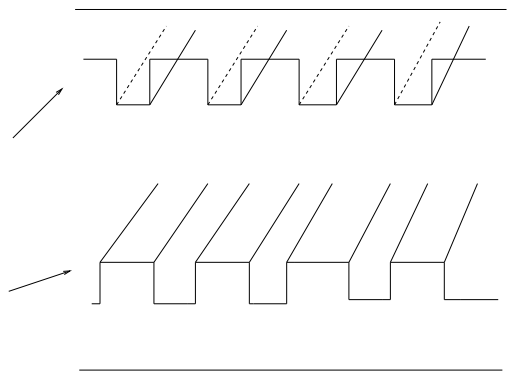}}
\caption{ Two parallel asymmetric arrays of optical waveguides that are
described by Eqs. (\ref{zeq1}) and (\ref{zeq2}), provided that parallel
beams propagate obliquely across both arrays. Arrows indicate misaligned
directions at which light is coupled into the arrays; inside them, both
propagation directions are identical.}
\label{figg1}
\end{figure}

There are at least two different physical realizations of the $\chi ^{(3)}$
model based on Eqs. (\ref{zeq1}) and (\ref{zeq2}). First, one may consider
two parallel arrays of nonlinear waveguides with different effective values 
$n^{(1)}$ and $n^{(2)}$ of the refractive index in them corresponding to a
given (oblique) direction of the light propagation. To this end, the
waveguides belonging to the two arrays may be fabricated from different
materials; alternatively, they may simply differ by the transverse size of
waveguiding cores, or by the refractive index of the filling between the
cores, see e.g.,  
Fig. \ref{figg1}. The difference in the effective
refractive index gives rise to the mismatch parameter $q$ in Eqs. 
(\ref{zeq1}) and (\ref{zeq2}). More importantly, it may 
also give rise to different
coefficients of the discrete diffraction. Indeed, the DM technique assumes
launching light into the array obliquely, the effective diffraction
coefficient in each array being \cite{Yaron2} 
\begin{equation}
D^{(1,2)}=2Cd^{2}\cos \left( k_{\bot }^{(1,2)}d\right) ,  \label{D}
\end{equation}
where $d$ is the spacing of both arrays, and $k_{\bot }^{(1,2)}$ are
transverse components of the two optical wave vectors. As it follows from
Eq. (\ref{D}), the diffraction coefficients are different if $k_{\bot
}^{(1)}\neq k_{\bot }^{(2)}$.

Despite the fact that $k_{\bot }^{(1)}$ and $k_{\bot }^{(2)}$ are assumed
different, we assume that the propagation directions of the light beams are
parallel in the two arrays, as a conspicuous walkoff (misalignment) between
them will easily destroy any coherent pattern. On the other hand, the light
coupled into both arrays has the same frequency, hence the absolute values
of the two wave vectors are related as follows: 
$k^{(1)}/k^{(2)}=n^{(1)}/n^{(2)}$, where $n^{(1,2)}$ are the above-mentioned
effective refractive indices. Combining the latter relation and the
classical refraction law, and taking into regard the condition that the
propagation directions are parallel inside the arrays, one readily arrives
at the conclusion that  
\begin{equation}
k_{\bot }^{(1)}/k_{\bot }^{(2)}=n^{(1)}/n^{(2)}.  \label{difference}
\end{equation}
Note that the two incidence angles $\theta ^{(1,2)}$ (at the interface
between the arrays and air) are related in a similar way, $\left( \sin
\theta ^{(1)}\right) /\left( \sin \theta ^{(2)}\right) =n^{(1)}/n^{(2)}$,
hence the incident beams (in air) must be {\em misaligned}, in order to be
aligned in the arrays.

Equation (\ref{D}) shows that there is a critical direction of the beam in
each array, corresponding to $k_{\bot }^{(1,2)}d=\pi /2$, at which the
effective diffraction coefficient changes its sign \cite{Yaron2}. Due to the
difference between $k_{\bot }^{(1)}$ and $k_{\bot }^{(2)}$, the critical
directions are different in the two arrays. Then, if the common propagation
direction in the arrays is chosen to be {\em between} the two critical
directions, Eq. (\ref{D}) gives different signs of the two diffractive
coefficients. Note that this interpretation of the model implies no XPM
coupling between the arrays, i.e., $\beta =0$ in Eqs. (\ref{zeq1}) and (\ref
{zeq2}).

An alternative realization is possible in a single array of {\em bimodal}
optical fibers, into which two parallel beams with orthogonal polarizations, 
$u$ and $v$, are launched obliquely. If the two polarizations are circular
ones, then $\beta =2$ in Eqs. (\ref{zeq1}) and (\ref{zeq2}), and the
asymmetry between the beams, which makes it possible to have different signs
of the coefficient (\ref{D}) for them, may be induced by birefringence,
which, in turn, can be easily generated by twist applied to the fibers \cite
{Agrawal}. The birefringence also gives rise to the mismatch $q$. As for the
linear mixing between the two polarizations, which is assumed in the model,
it can be easily induced if the fibers are, additionally, slightly deformed,
having an elliptic cross section \cite{Agrawal}. If the two polarizations
are linear, then the birefringence is induced by the elliptic deformation,
and the linear mixing is induced by the twist, the XPM coefficient being 
$2/3$ in this case (assuming that, as usual, the birefringence makes it
possible to neglect four-wave mixing nonlinear terms \cite{Agrawal}).

It is interesting to note that the discrete model based on Eqs. (\ref{zeq1})
and (\ref{zeq2}) with $\kappa =0$ is exactly tantamount to a formal
discretization of the above-mentioned continuum model which was introduced
in Ref. \cite{Dave} to describe a dual-core optical fiber with opposite
signs of dispersion in the cores. Another quite noteworthy feature of the
present model is that, if $\beta =2$, it turns out to be formally equivalent
to a discretization of the standard Bragg-grating model (\ref{BG}), which is
produced by replacing $\Psi _{x}\rightarrow \left( \Psi _{n+1}-\Psi
_{n-1}\right) /2$ and $\Phi _{x}\rightarrow \left( \Phi _{n+1}-\Phi
_{n-1}\right) /2$. Indeed, making the substitution (``staggering
transformation'') 
\begin{equation}
\Psi _{n}\equiv i^{n}\phi _{n},\Phi _{n}\equiv i^{n}\psi _{n},
\label{stagger}
\end{equation}
one concludes that the discrete version of Eqs. (\ref{BG}) takes precisely
the form of Eqs. (\ref{zeq1}) and (\ref{zeq2}) with $\delta
=1,\,q=2 C,\,\kappa =1$, and $\beta =2$.

\subsection{The linear spectrum}

Before proceeding to the presentation of numerical results for solitons
found in the system of Eqs. (\ref{zeq1}) and (\ref{zeq2}), it is relevant to
understand at which values of the propagation constant $\Lambda $ (spatial
frequency) solitons with exponentially decaying tails may exist in this
model. There are two regions in which they may be found. Firstly, inside the
gap of the system's linear spectrum one may find 
{\em discrete gap solitons}, i.e., counterparts of the GSs found 
in the continuum version of the model
in Ref. \cite{Dave}. Secondly, solitons specific to the discrete model may
be found {\em above} the phonon band. To analyze these possibilities, an
asymptotic expression for the tail, 
\begin{equation}
\psi _{n},\phi _{n}\sim \exp \left( i\Lambda t-\lambda \left| n\right|
\right)  \label{tail}
\end{equation}
is to be substituted into the linearized version of Eqs. (\ref{zeq1}) and 
(\ref{zeq2}).

Investigating the possibility of the existence of solitons above the phonon
gap, it is sufficient to focus on the particular case $\delta =1$ and $q=0$,
when the system's spectrum takes a simple form (we have also considered more
general cases with positive $\delta $ different from $1$ and $q\neq 0$,
concluding that they do not yield anything essentially different from this
case). The final result, produced by a straightforward algebra, is that
solitons are possible in the region 
\begin{equation}
\Lambda ^{2}>\Lambda _{{\rm edge}}^{2}\equiv 16C^{2}+\kappa ^{2}\text{,}
\label{above_band}
\end{equation}
\newline
$\pm \Lambda _{{\rm edge}}$ being edges of the phonon band. In what follows
below, we will assume $\Lambda >0$, as in this case positive and negative
values of $\Lambda $ are equivalent.

To understand the possibility of the existence of the discrete GSs, we,
first, set $\delta =1$ as above, but keep the mismatch $q$ as an arbitrary
parameter. Then, the gap is easily found to be 
\begin{equation}
\Lambda ^{2}<\Lambda _{{\rm gap}}^{2}\equiv \left\{ 
\begin{array}{cc}
q^{2}+\kappa ^{2} & {\rm if}\,\,$q<0$, \\ 
\kappa ^{2} & {\rm if\,}\,0\leq q\leq 4C, \\ 
\left( q-4C\right) ^{2}+\kappa ^{2} & {\rm if\,}\,$q>4C$\,
\end{array}
\right.  \label{gap}
\end{equation}
(recall that, by definition, $C>0$). An essential role of the mismatch
parameter is that it makes the gap broader if it is negative.

In the more general case, $\delta \neq 1$, two different layers can be
identified in the gap, similar to what was found in the continuum limit \cite
{Dave}. For instance, if $q=0$, the {\it inner} and {\it outer} layers are 
\begin{equation}
0<\Lambda ^{2}<\frac{4\delta }{\left( \delta +1\right) ^{2}}\kappa ^{2},\, 
{\rm and}\,\,\frac{4\delta }{\left( \delta +1\right) ^{2}}\kappa
^{2}<\Lambda ^{2}<\kappa ^{2}  \label{innerouter}
\end{equation}
(in the case $\delta =1$, the outer layer disappears). The difference
between the layers is the same as in the continuum limit \cite{Dave}: in the
outer layer, solitons, if any, have monotonically decaying tails, i.e., real 
$\lambda $ in Eq. (\ref{tail}), while in the inner layer $\lambda $ is
complex, and, accordingly, soliton tails are expected to decay with
oscillations.

\subsection{The structure of the work}

The rest of the paper is organized as follows. In section II, we display
results for solitons found above the phonon band, i.e., in the region (\ref
{above_band}). The evolution of the solitons is monitored, starting from the
anti-continuum (AC) limit $C=0$, and gradually increasing $C$. Any branch of
soliton solutions in this region must disappear, approaching the continuum
limit. Indeed, as the radiation band (frequently called ``phonon band'',
referring to linear phonon modes in the lattice dynamics) becomes infinitely
broad in this limit, see Eq. (\ref{above_band}), the solution branch with 
$\Lambda ={\rm const}$ will crash hitting the swelling phonon band. However,
in many cases the soliton of this type is found to remain stable {\em as
long as it exists}, so it may be easily observed experimentally in the
optical array.

In section III we present results for solitons existing inside the gap. In
the outer layer [which is defined as per Eq. (\ref{innerouter}), provided
that $\delta \neq 1$], we were able to find only solitons of an ``antidark''
type, that sat on top of a nonvanishing background. However, in the inner
layer [recall it occupies the entire gap in the case $\delta =1$, according
to Eq. (\ref{innerouter})], true solitons are easily found (in accord with
the prediction, their tails decay with oscillations). In the case $q=0$,
these solutions appear as stable ones in the AC limit, get destabilized at
some finite critical value of $C$, and continue, as unstable solutions, all
the way up to the continuum limit, never disappearing. It is quite
interesting that sufficiently large negative mismatch {\em strongly extends}
the stability range for these solitons.

As was mentioned above, the $\chi ^{(3)}$ model based on Eqs. (\ref{zeq1})
and (\ref{zeq2}) may be considered as a discretization of the standard
gap-soliton system (\ref{BG}). In this connection, it is natural to search
for discrete counterparts of the usual GSs in the latter system. However,
the discrete GSs found in section III do not have any counterpart in the
continuum system (\ref{BG}), as the staggering transformation 
(\ref{stagger}) makes direct transition from the discrete 
equations (\ref{zeq1}) and 
(\ref{zeq2}) to the continuum system (\ref{BG}) impossible. 
At the end of section
III, we specially consider discrete solitons which are directly related to
GSs in the system (\ref{BG}). We find that such solitons exist indeed at all
the values of $C$, their drastic difference from those found in sections II
and III is that they are essentially complex solutions to the stationary
version of Eqs. (\ref{zeq1}) and (\ref{zeq2}). At all finite values of $C$,
they are unstable, but the instability asymptotically vanishes in the AC and
continuum limits, $C\rightarrow 0$ and $C\rightarrow \infty $.

In section IV, we briefly consider the SHG model (\ref{jeq12}), 
(\ref{jeq13}). Solitons are found in this model too, and their stability is
investigated. When the solitons are linearly unstable, the development of
their instability is examined (in all the sections II, III, and IV) by means
of direct numerical simulations, which show that the instability may
initiate a transition to a localized breather, or to lattice turbulence, or,
sometimes, complete decay of the soliton into lattice phonon waves.

\section{Solitons above the phonon band}

\subsection{General considerations}

Stationary solutions to Eqs. (\ref{zeq1})-(\ref{zeq2}) are sought for the
form 
\begin{equation}
\psi _{n}=e^{i\Lambda t}u_{n},\,\phi _{n}=e^{i\Lambda t}v_{n}\,,
\label{zeq4}
\end{equation}
where $\Lambda $ is the propagation constant defined above. In figures
displayed below, the stationary solutions will be characterized by the norms
of their two components, 
\begin{equation}
P_{u}^{2}\equiv \sum_{n=-\infty }^{+\infty }u_{n}^{2},\,\,P_{v}^{2}\equiv
\sum_{n=-\infty }^{+\infty }v_{n}^{2}\,.  \label{P}
\end{equation}
Once such solutions are numerically identified by means of a Newton-type
numerical scheme, we then proceed to investigate their stability, assuming
that the solution is perturbed as follows: 
\begin{eqnarray}
\psi _{n} &=&\left[ u_{n}+\epsilon a_{n}\exp (i\omega t)+\epsilon b_{n}\exp
(-i\omega ^{\star }t)\right] \exp (i\Lambda t),  \label{zeq5} \\
\phi _{n} &=&\left[ v_{n}+\epsilon c_{n}\exp (i\omega t)+d_{n}\exp (-i\omega
^{\star }t)\right] \,\exp (i\Lambda t),  \label{zeq6}
\end{eqnarray}
where $\epsilon $ is an infinitesimal amplitude of the perturbation, and 
$\omega $ is the eigenvalue corresponding to the linear (in)stability mode.
The set of the resulting linearized equations for the perturbations 
$\{a,b^{\star },c,d^{\star };\omega \}$ is subsequently solved as an
eigenvalue problem. This is done by using standard numerical linear algebra
subroutines built into  mathematical software packages \cite{Matlab}.
If all the eigenvalues $\omega $ are purely real, the
solution is marginally stable; on the contrary, the presence of a nonzero
imaginary part of $\omega $ indicates that the soliton is unstable.
When the solutions were unstable, their dynamical evolution was followed
by means of fourth-order Runge-Kutta numerical integrators, to identify
the development and outcome of the corresponding instabilities.

In what follows below, we describe different classes of soliton solutions,
which are generated, in the AC limit, by expressions with different
symmetries. Still another class of solitons, which carries over into the usual
GSs in the continuum system (\ref{BG}), will be considered in the next
section.

\subsection{Solution families which are symmetric in the anti-continuum 
limit}

\begin{figure}[tbp]
\epsfxsize=8.5cm
\centerline{\epsffile{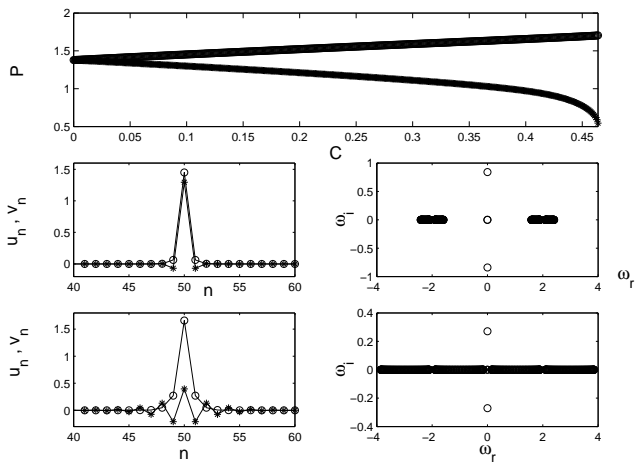}}
\epsfxsize=8.5cm
\centerline{\epsffile{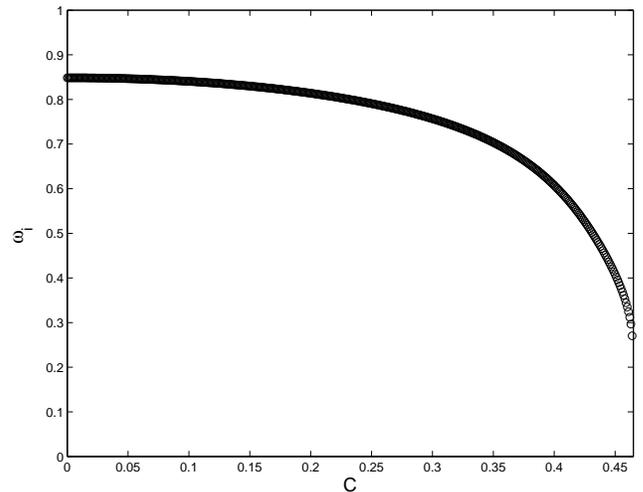}}
\caption{ The top panel shows the norms $P_{u}$ (lower curve) and $P_{v}$
(upper curve) of the two components of the soliton solution vs. $C$, up to
the point where the branch terminates. The next set of panels shows two
examples of the solution at $C=0.1$ (the upper row) and $C=0.464$ (just near
the termination point of the branch; the lower row), together with the
spectral planes of the corresponding linear stability eigenvalues (the
vertical and horizontal coordinates in the plane correspond to the imaginary
and real parts of $\protect\omega$). The profiles of the $u_n$ and $v_n$
components are shown, respectively, by circles and stars. These solutions
are always unstable. The bottom panel shows the imaginary part of the single
unstable eigenfrequency vs. $C$. }
\label{jfig1}
\end{figure}

As it was said above, in this section we set $\delta =1$ and $q=0$, since
comparison with more general numerically found results has demonstrated that
this case adequately represents the general situation, as concerns the
existence and stability of solitons. Figure \ref{jfig1} shows a family of
soliton solutions found for $\kappa =0.1$, $\Lambda =2$ and $\beta =0$, as a
function of the coupling constant $C$. In this case, the family starts, in
the AC\ limit ($C=0$), with a solution that consists of a symmetric
excitation localized at a single lattice site $n_{0}$, with 
\begin{equation}
u_{n_{0}}=v_{n_{0}}=\pm \sqrt{\frac{\Lambda -\kappa }{1+\beta }}\,,
\label{AC1}
\end{equation}
and terminates at finite $C$. Figure \ref{jfig1} demonstrates that this
branch is always unstable. The termination of the branch happens when it
comes close to the phonon band, that swells with the increase of $C$. The
branch terminates at $C=0.464$, when the upper edge of the band is at 
$\Lambda _{{\rm edge}}=\sqrt{\kappa ^{2}+16C^{2}}\approx 1.859$, according to
Eq. (\ref{above_band}). This value is still smaller than the fixed value of
the soliton's propagation constant, $\Lambda =2$, for which the soliton
branch is displayed in Fig. \ref{jfig1}. The branch, if it could be
continued, would crash into the upper edge of the phonon band at $C=0.499$.
The slightly premature termination of this soliton family is a consequence
of the nonlinear character of the solutions, as the above prediction for the
termination point was based on the linear approximation.

\begin{figure}[tbp]
\epsfxsize=8.5cm
\centerline{\epsffile{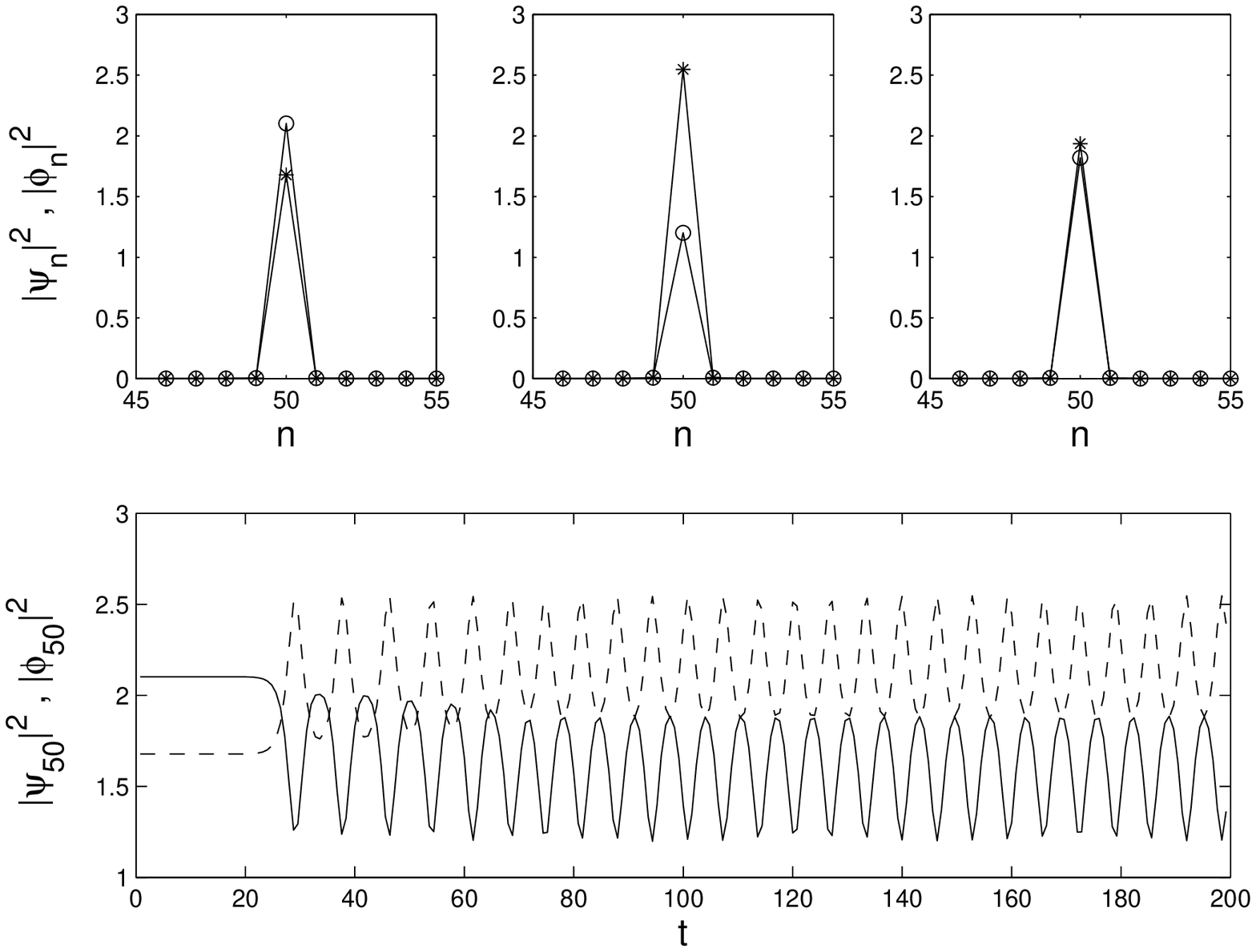}}
\caption{Evolution of the unstable soliton from Fig. \ref{jfig1}) in the
case $C=0.1$, $\protect\beta=0$, and $\protect\kappa=0.1$. The top panel
shows the fields' spatial profiles (the circles correspond to 
$|\protect\psi_n|^2$, and the stars to $|\protect\phi_n|^2$) 
for $t=4$ (left panel), 
$t=192 $ (middle panel) and $t=196$ (right panel). The first profile is
nearly identical to the initial condition, while the other two were chosen
close to points where the oscillating amplitude of the resultant breather
attains its maximum and minimum. The bottom panel shows the field evolution
at the central lattice site ($n=50$), clearly demonstrating the breathing
nature of the established state. The solid and dashed lines are,
respecively, $|\protect\psi_{50}|^2$ and $|\protect\phi_{50}|^2$. In this
case, the instability growth rate of the initial soliton is $\approx 0.8$;
in view of this large value, it was not necessary to add any initial
perturbation to trigger the instability.}
\label{newfig1}
\end{figure}

An example of the development of the instability of this solution, as found
from direct simulations of the full equations (\ref{zeq1}) and (\ref{zeq2}),
is given in Fig. \ref{newfig1} for $C=0.1$. It is clearly seen that the
unstable soliton turns into a stable breather.

\begin{figure}[tbp]
\epsfxsize=8.5cm
\centerline{\epsffile{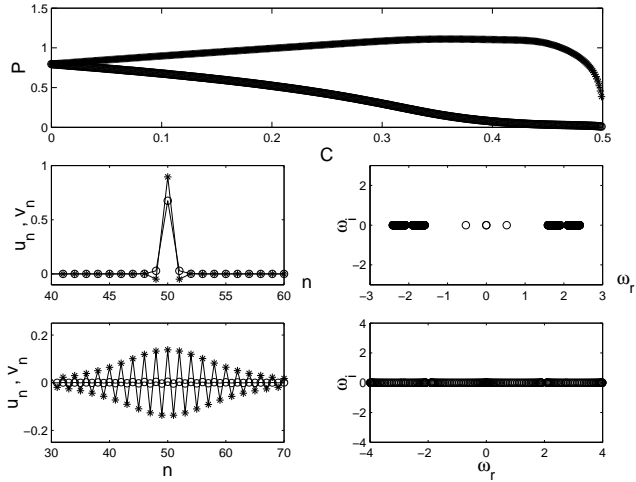}}
\caption{The same as in Fig. \ref{jfig1}, but for the case 
$\protect\beta= 2$. The middle and lower panels display examples 
of the soliton solutions at 
$C=0.1$ and $C=0.499$, respectively. In the top panel, the upper and lower
curves now correspond to the $v_n$ and $u_n$ components, i.e., opposite to
the case shown in Fig. 1. Notice that this branch is {\it always} stable
until it terminates, therefore the figure does not contain a counterpart of
the dependence shown in the bottom panel of Fig. \ref{jfig1}.}
\label{jfig2}
\end{figure}

On the contrary, in the presence of XPM with the physically relevant value
of $\beta =2$, a similar solution branch, found for the same values $\kappa
=0.1$ and $\Lambda =2$, is {\em stable} for all $C$, until it terminates at 
$C=0.499$. Note that, at this point, the upper edge (\ref{above_band}) of the
phonon band is $\Lambda _{{\rm edge}}=1.999$, which is extremely close to 
$\Lambda =2$, i.e., the termination of the solution family is indeed
accounted for by its crash into the swelling phonon band. Details of this
stable branch are shown in Fig. \ref{jfig2}.

Direct simulations of this solution have corroborated its stability (details
are not shown here). In fact, in {\em all} the cases when solitons are found
to be stable in terms of the linearization eigenvalues (see other cases
below), direct simulations fully confirm their dynamical stability.

\subsection{Solution families which are anti-symmetric in the anti-continuum
limit}

\begin{figure}[tbp]
\epsfxsize=8.5cm
\centerline{\epsffile{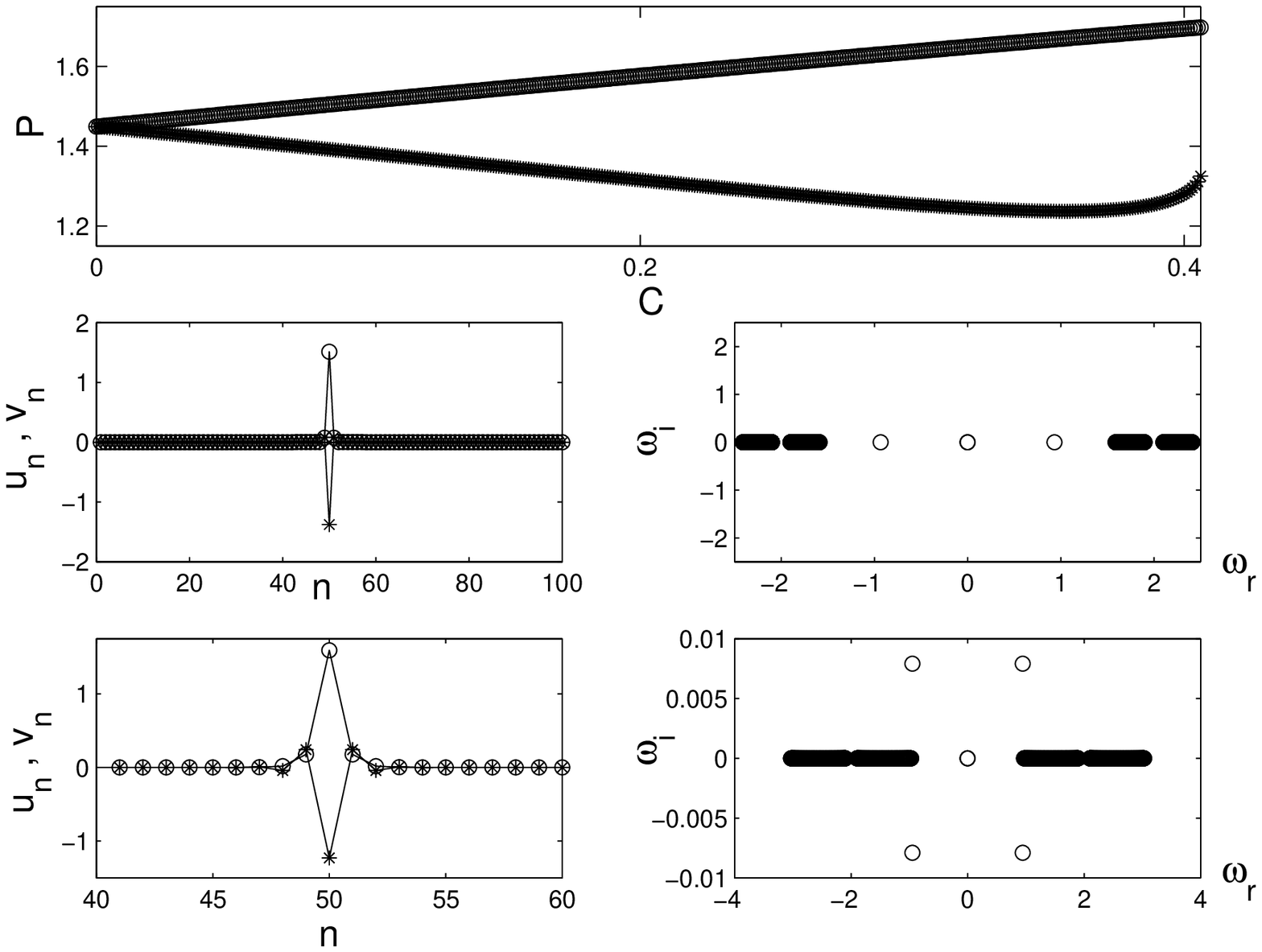}}
\epsfxsize=8.5cm
\centerline{\epsffile{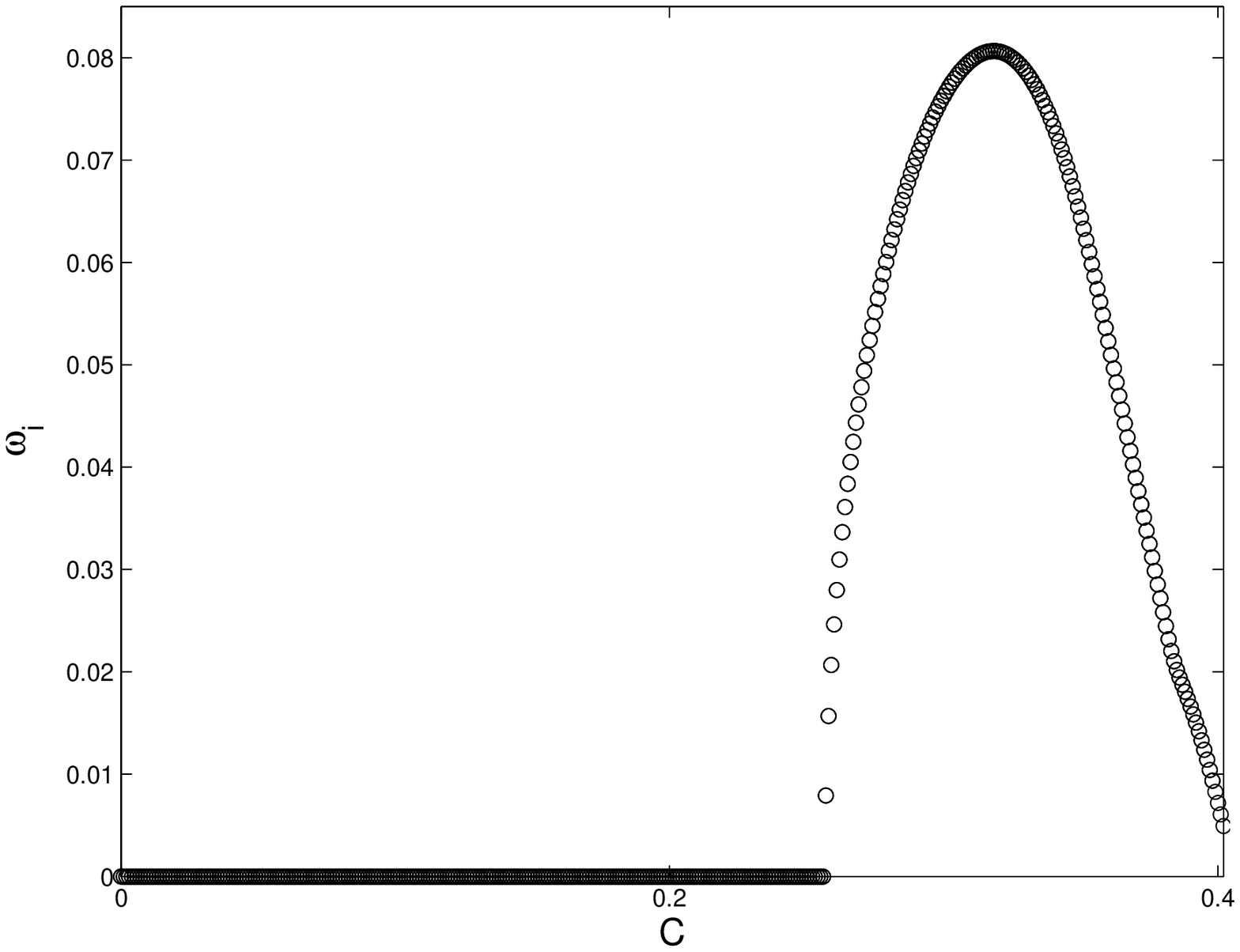}}
\epsfxsize=8.5cm
\centerline{\epsffile{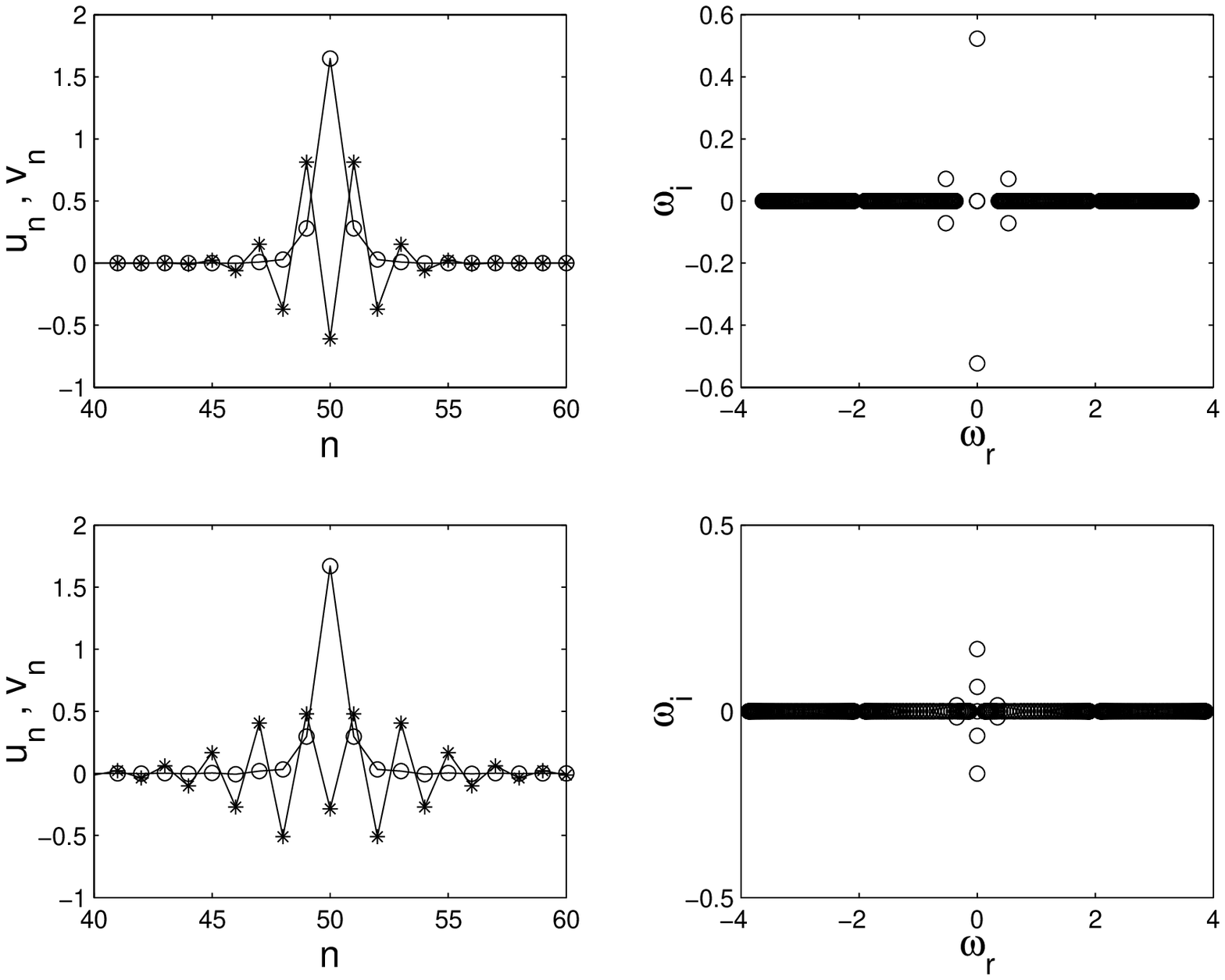}}
\caption{The top panel shows the branch of solutions starting from the
expression (\ref{AC2}) in the AC limit. Two particular examples are shown
for $C=0.1$ (a stable solution, the upper row) and $C=0.257$ (at the onset
of the oscillatory instability, the lower row). The circles and stars again
denote the $u_n$ and $v_n$ components, respectively. The panel beneath this
displays the instability growth rate, ${\rm Im}\,\protect\omega$, vs. $C$.
Finally, the bottom panels show the profiles and linear stability
eigenvalues for another solution, found beyond the termination point of this
branch at $C=0.407$. Two examples of the new solution are given for $C=0.407$
(the upper row) and $C=0.467$ (the lower row). These two points are very
close to the beginning and termination of the new branch).}
\label{jfig3}
\end{figure}

Another branch of solutions is initiated, in the AC limit, by an
anti-symmetric excitation localized at a single lattice site, cf. Eq. (\ref
{AC2}): 
\begin{equation}
u_{n_{0}}=-v_{n_{0}}=\pm \sqrt{\frac{\Lambda +\kappa }{1+\beta }}.
\label{AC2}
\end{equation}
The solution belonging to this branch is shown in Fig. \ref{jfig3} for the
same values of parameters as in Fig. \ref{jfig1}, i.e., $\delta =1$, $\kappa
=0.1$, $\Lambda =2$, and $\beta =0$. With the increase of $C$, this branch
picks up an oscillatory instability at $C\approx 0.257$, and terminates at 
$C\approx 0.407$. Unlike the solutions displayed above, the termination of
this branch occurs {\em not} through its crash into the phonon band, but via
a saddle-node bifurcation. The latter bifurcation implies a collision with
another branch of solutions. That additional branch (which is strongly
unstable) was found but is not shown here.

In fact, the numerical algorithm is able to capture other solutions
(unstable ones) past the point $C\approx 0.407$ at which the present
solution terminates. The newly found solutions are shown in the bottom part
of Fig. \ref{jfig3}. However, the new family cannot be continued beyond 
$C=0.467$ [cf. the termination point $C=0.464$ for the solutions initiated in
the AC limit by the expression (\ref{AC1})].

\begin{figure}[tbp]
\epsfxsize=8.5cm
\centerline{\epsffile{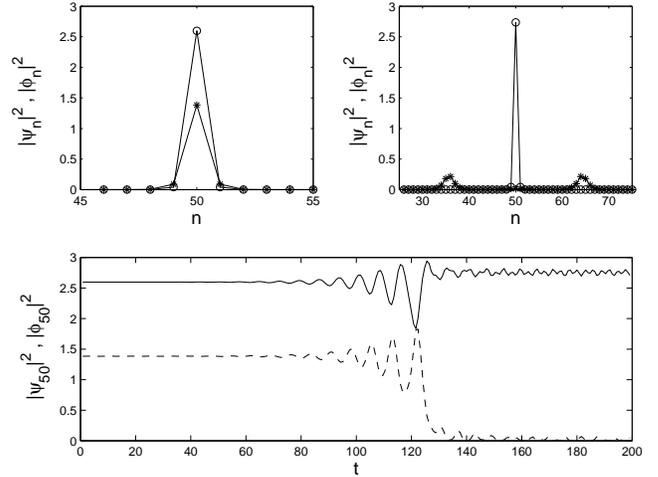}}
\caption{Dynamical development of the oscillatory instability of the
anti-symmetric solution for $C=0.3$, $\protect\beta=0$ and $\protect\kappa
=0.1$. The meaning of the symbols is as in Fig. \ref{newfig1}. The top left
and right panels show the field configuration at $t=4$ and $t=196$,
respectively. The bottom panel once again shows the field evolution at the
central site.}
\label{newfig2}
\end{figure}

The development of the oscillatory instability of the solution shown in Fig. 
\ref{jfig3} was also studied in direct simulations. It leads to onset of a
state where one component of the soliton is fully destroyed [it cannot
completely disappear, due to the presence of the linear couplings in Eqs. 
(\ref{zeq1}) and (\ref{zeq2}), but it is reduced to a level of small random
noise]. An example of this is given in Fig. \ref{newfig2} for 
$C=0.3$, $\beta =0$ and $\kappa =0.1$. The instability 
(with the initial growth rate 
$0.07$ in this case) develops after $t\approx 60$, destroying one component
of the soliton in favor of further growth of the other one. In this case, a
uniformly distributed noise perturbation of an amplitude $10^{-4}$ was added
to accelerate the onset of the instability, as the initial instability is
very weak (which implies that the unstable soliton may be observed in
experiment).

\begin{figure}[tbp]
\epsfxsize=8.5cm
\centerline{\epsffile{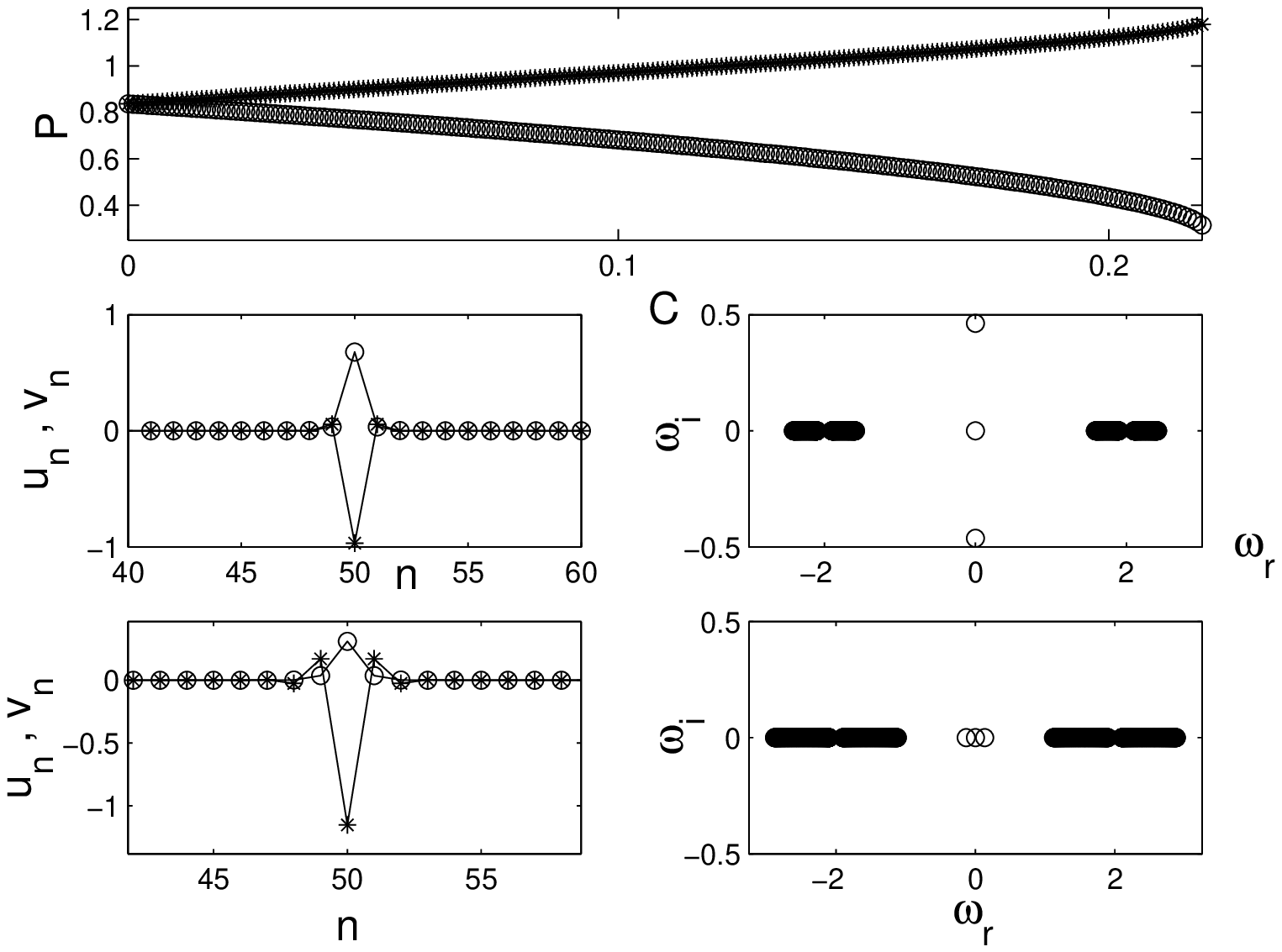}}
\epsfxsize=8.5cm
\centerline{\epsffile{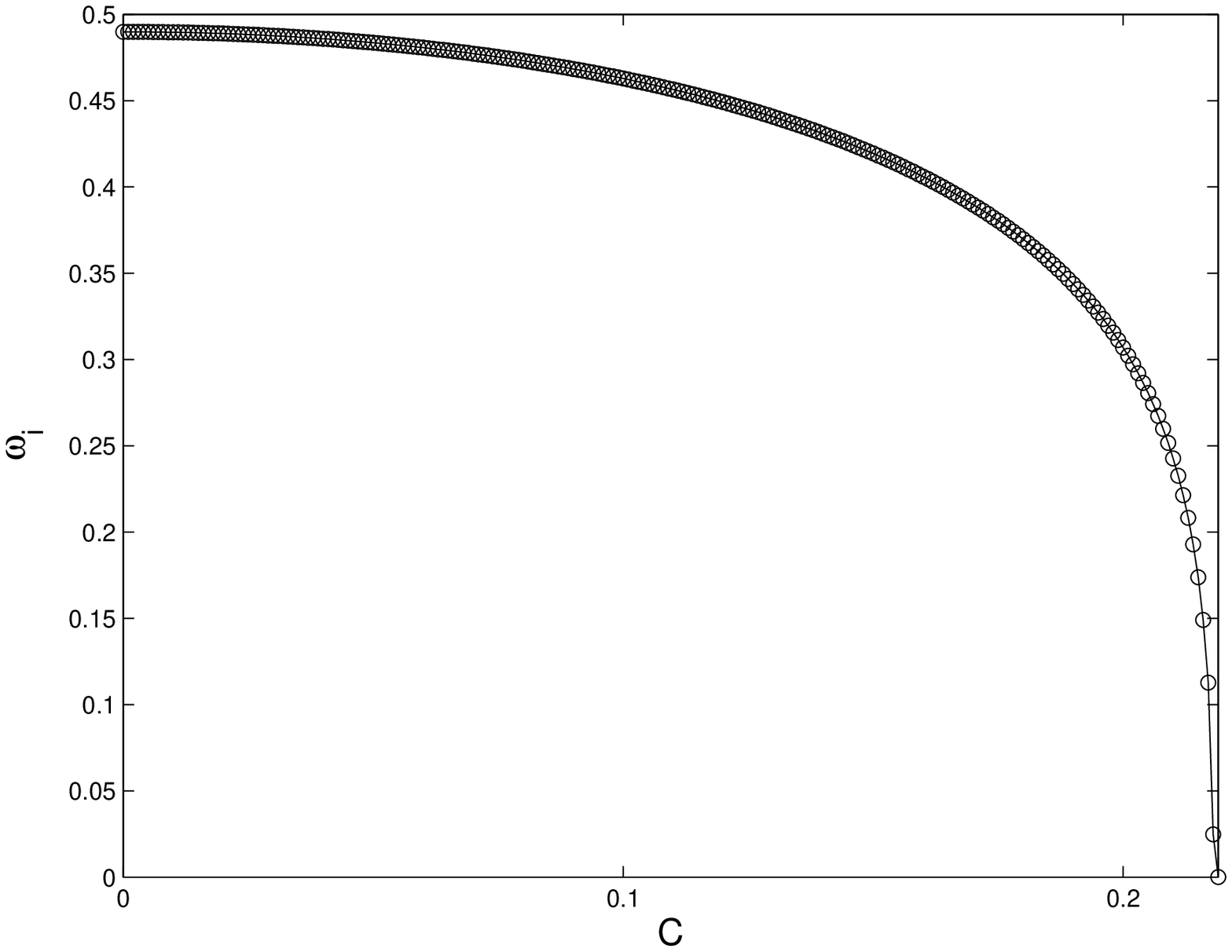}}
\epsfxsize=8.5cm
\centerline{\epsffile{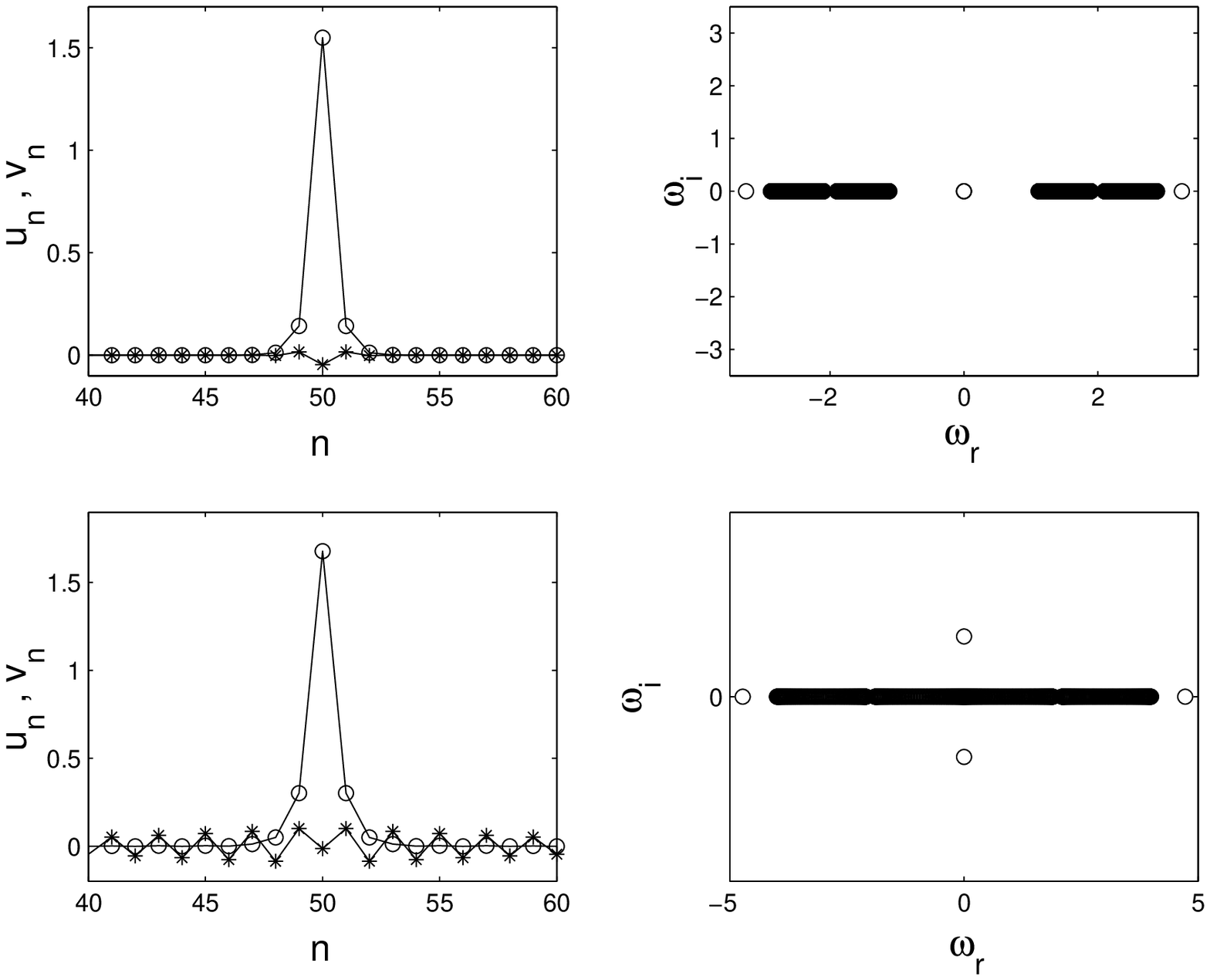}}
\caption{ The same as in Fig. \ref{jfig3}, but for $\protect\beta=2$. This
branch is {\it always} unstable (as is shown by the middle plot
demonstrating the instability growth rate vs. $C$) in its range of
existence, $0<C<0.219$. Examples of the solution displayed in the upper part
of the figure are given for $C=0.1$ and $C=0.219$. The lower part shows the
new solution family found past the termination point of the unstable branch.
Example of the new solutions are given for $C=0.22$ (stable, the upper row)
and $C=0.498$ (just prior to the termination of the new family, the lower
row). The instability of this branch sets in at $C \approx 0.494$, i.e.,
very close to the termination point. }
\label{jfig4}
\end{figure}

A counterpart of the solution from Fig. \ref{jfig3}, but with $\beta =2$,
rather than $\beta =0$, is shown in Fig. \ref{jfig4}. This branch is always
unstable (i.e., in the case of the solutions starting from the
anti-symmetric expression in the AC\ limit, the XPM nonlinearity
destabilizes the solitons, while in the case of the branch that was
initiated by the symmetric expression in the AC limit, the same XPM
nonlinearity was stabilizing). It terminates at $C\approx 0.219$, again
through a saddle-node bifurcation. As in the previous case, a new family of
solutions can be captured by the numerical algorithm past the termination
point. The new family is found for $0.22<C<0.498$, and it is also shown in
Fig. \ref{jfig4}. Comparing the value $\Lambda _{{\rm edge}}=1.995$ given by
Eq. (\ref{above_band}) in this case with the actual value $\Lambda =2$ of
the soliton's propagation constant, we conclude that the termination of the
latter branch is caused by its collision with the phonon band. Notice also
that the latter branch becomes unstable only very close to its termination
point, at $C>0.494$.

\begin{figure}[tbp]
\caption{The development of the instability accounted for by the imaginary
eigenfrequency (with the growth rate $\approx 0.45$) of the anti-symmetric
branch, in the case of $C=0.1$, $\protect\kappa=0.1$. The top panels pertain
to $t=4$ (left), $t=100$ (middle) and $t=120$ (right). The latter two have
again been chosen close to the points where the oscillating amplitude of the
resultant breather attains its maximum and minimum, respectively. The
instability sets in around $t \approx 40$; no external perturbation was
added to the initial condition in this case. }
\epsfxsize=8.5cm
\centerline{\epsffile{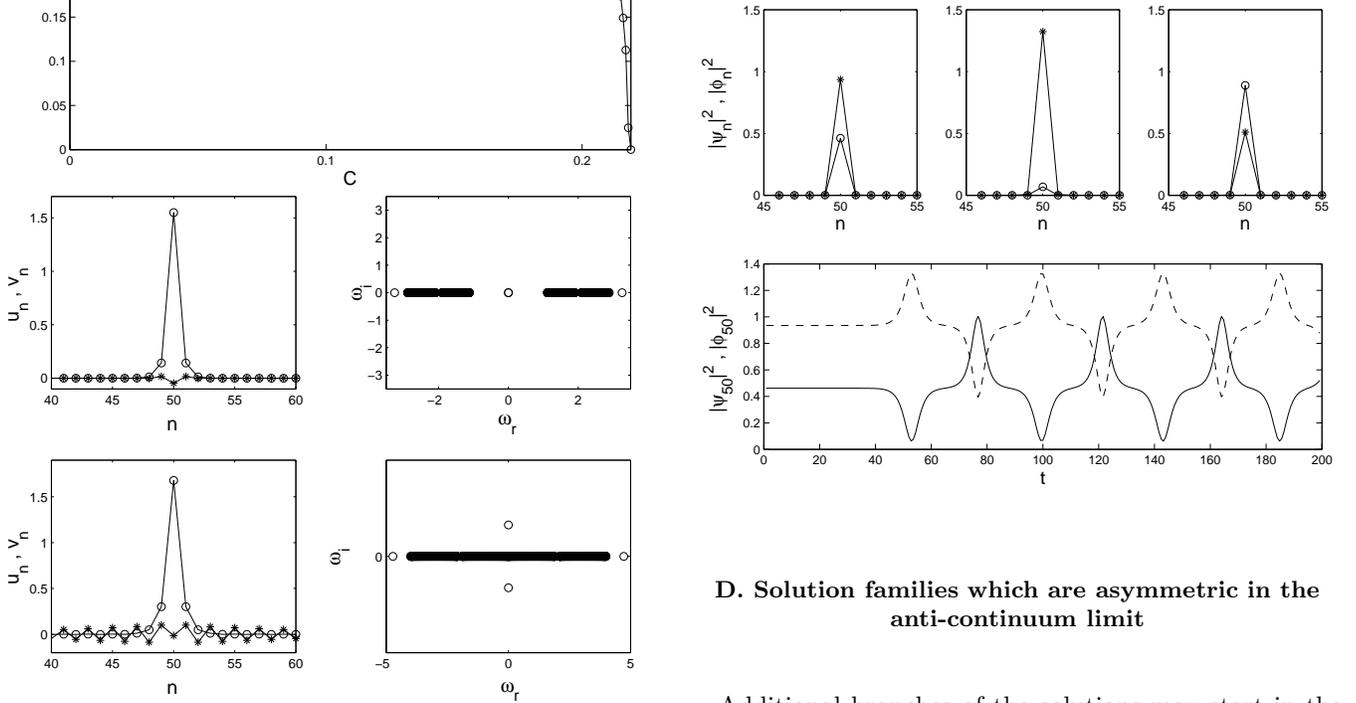}}
\label{newfig3}
\end{figure}

In the case of $\beta =2$, direct simulations show that the instability of
the anti-symmetric branch gives rise to rearrangement of the solution into a
very regular breather shown in Fig. \ref{newfig3} for $C=0.1$ and $\kappa
=0.1$.

\subsection{Solution families which are asymmetric in the anti-continuum
limit}

\begin{figure}[tbp]
\epsfxsize=8.5cm
\centerline{\epsffile{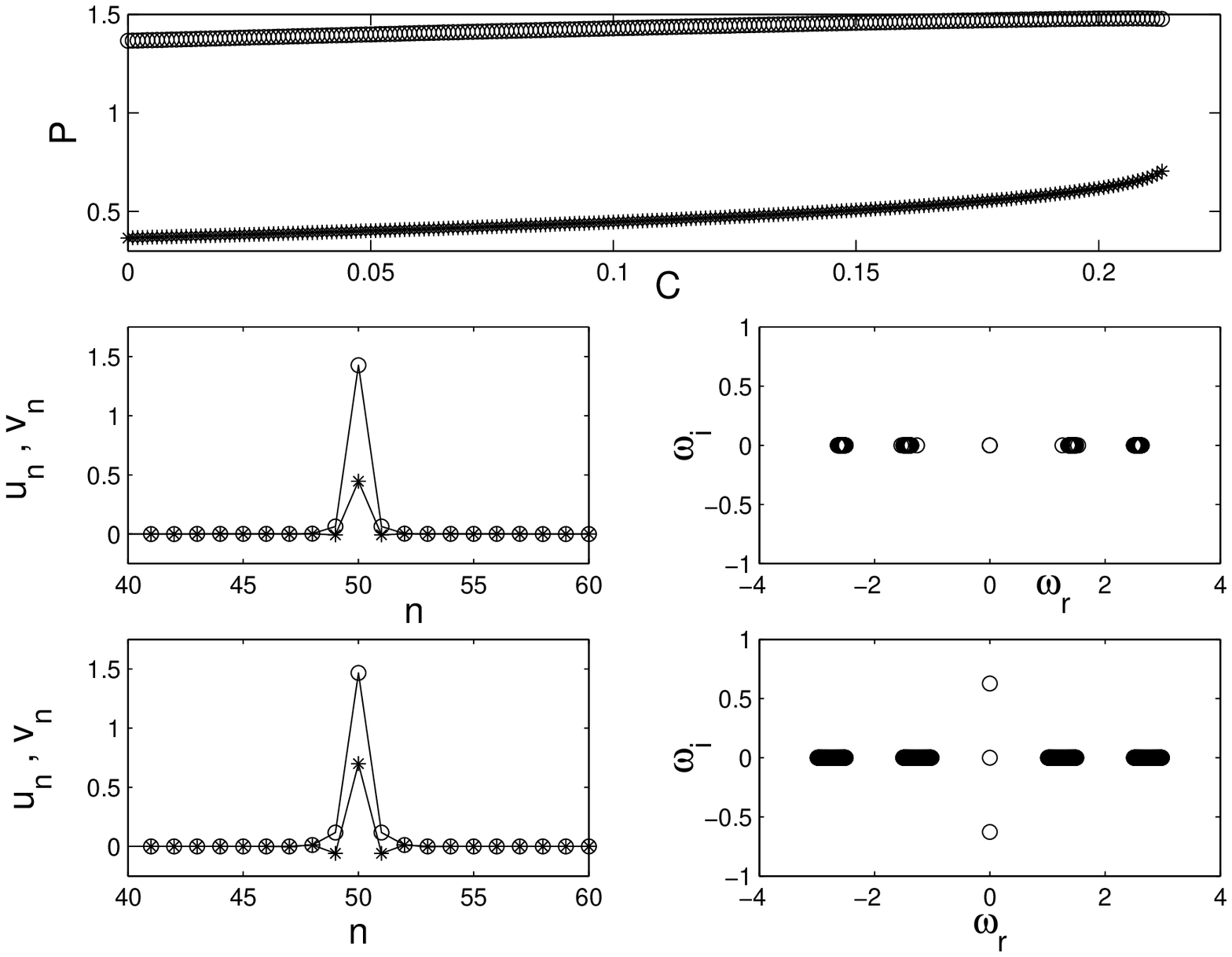}}
\epsfxsize=8.5cm
\centerline{\epsffile{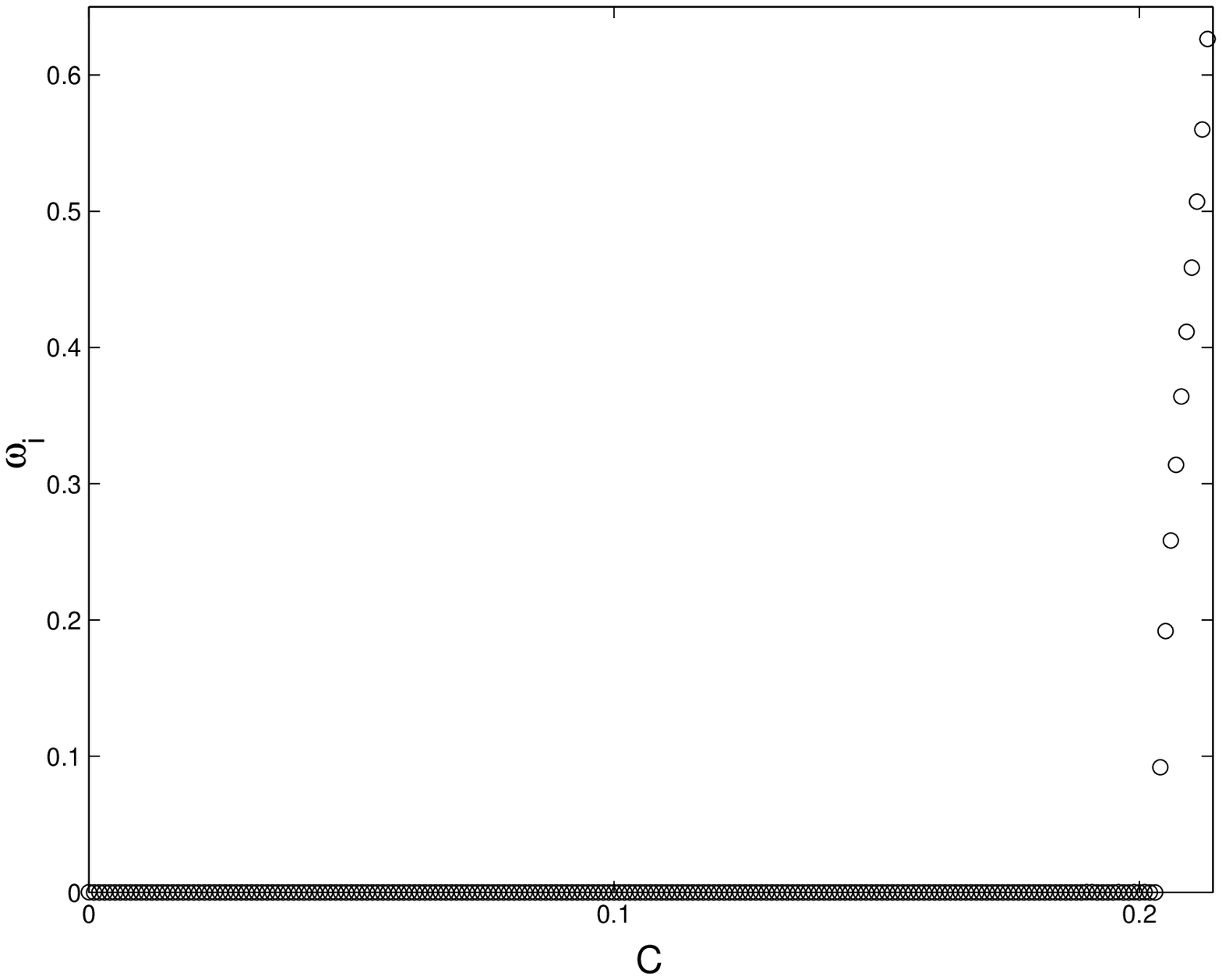}}
\caption{ The solution branch generated, in the AC limit, by the asymmetric
expression (\ref{jeq9})-(\ref{jeq10}) with the upper sign. The notation is
the same as in Fig. \ref{jfig1}. Two examples of the solution are shown for 
$C=0.1$ and $C=0.213$. The most unstable eigenvalue is shown, vs. $C$, in the
bottom panel. The instability sets in at $C\approx 0.204$, and the branch
terminates at $C\approx 0.213$. }
\label{jfig5}
\end{figure}

Additional branches of the solutions may start in the AC limit from {\em
asymmetric} configurations, provided that $\Lambda $ is still larger, namely
for $\Lambda >2\kappa $. In particular, such an extra branch can be
initiated by the following AC-limit solution excited at a single site 
$n=n_{0}$ (here, $\beta =0$), cf. Eqs. (\ref{AC1}) and (\ref{AC2}): 
\begin{eqnarray}
u_{n_{0}}^{2} &=&{\frac{1}{2}\left[ \Lambda \pm \sqrt{\Lambda ^{2}-4\kappa
^{2}}\right] },  \label{jeq9} \\
v_{n_{0}} &=&\kappa ^{-1}(\Lambda u_{n_{0}}-u_{n_{0}}^{3})\,.  \label{jeq10}
\end{eqnarray}
An example of this solution for the upper sign in Eq. (\ref{jeq9}) is shown,
for $\Lambda =2$, $\kappa =0.5$ and $\delta =1$, in Fig. \ref{jfig5}. Such
asymmetric branches may be stable for sufficiently weak coupling (in this
case, for $C<0.204$), but they eventually become unstable, and disappear
soon thereafter (at $C\approx 0.213$, in this case).

The evolution of the instability (for $C>0.204$) for this asymmetric branch
is strongly reminiscent of that shown in Fig. \ref{newfig1}, resulting in a
persistent breathing state.

\begin{figure}[tbp]
\epsfxsize=8.5cm
\centerline{\epsffile{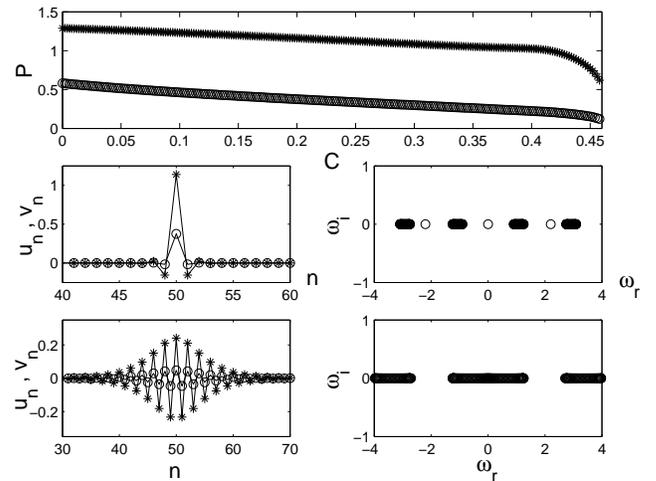}}
\caption{ The same as in Fig. \ref{jfig5}, but generated by the expression 
(\ref{jeq9}) with the lower sign. This solution is always stable until it
terminates at $C \approx 0.458$. Examples of the solution for $C=0.2$ and 
$C=0.458$ (the latter case is chosen just prior to the termination of the
branch) are shown, as usual, by means of their profiles and linear stability
eigenfrequencies.}
\label{jfig5a}
\end{figure}

The branch that commences from the AC expression (\ref{jeq9}) with the lower
sign is shown for $\Lambda =2$, $\kappa =0.75$ and $\delta =1$ in Fig. \ref
{jfig5a}. The branch remains {\em stable} as long as it exists, i.e., for 
$C<0.46$. At this point, it disappears colliding with the phonon band, whose
upper edge is located, according to Eq. (\ref{above_band}), at 
$\Lambda _{{\rm edge}}\approx 1.987$, which is very close to the family's fixed
propagation constant, $\Lambda =2$.

\section{Gap solitons}

\subsection{Solitons in the inner layer of the gap}

All the solutions that were examined in the previous section had their
propagation constant {\it above} the upper edge of the phonon spectrum.
Another issue of obvious interest is to study possible gap solitons (GSs),
whose propagation constant is located inside the gap (\ref{gap}), i.e., {\it
below} the lower edge of the phonon band. Unlike the solitons found above
the band, GSs may persist up to the continuum limit.

\begin{figure}[tbp]
\epsfxsize=8.5cm
\centerline{\epsffile{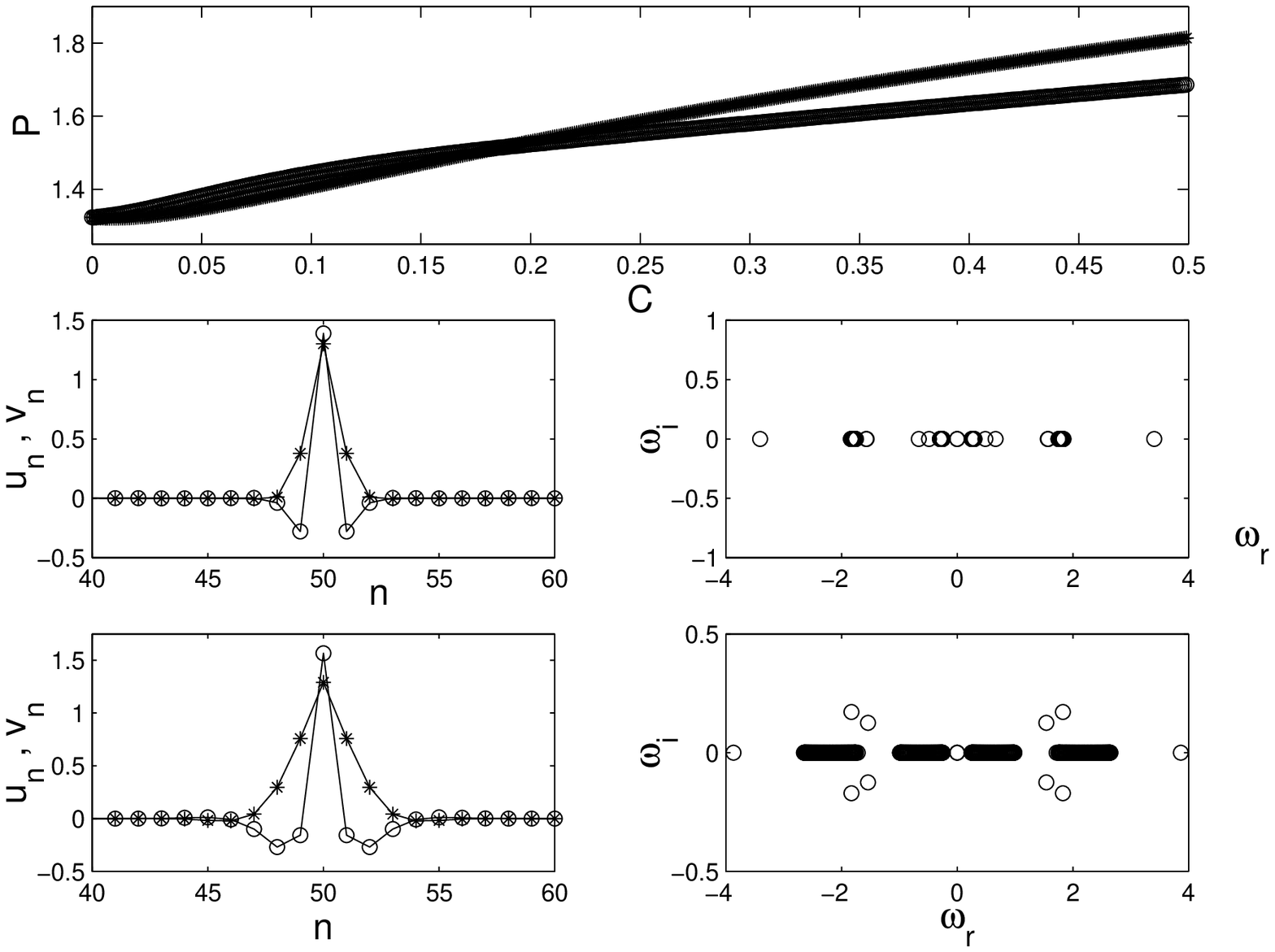}}
\epsfxsize=8.5cm
\centerline{\epsffile{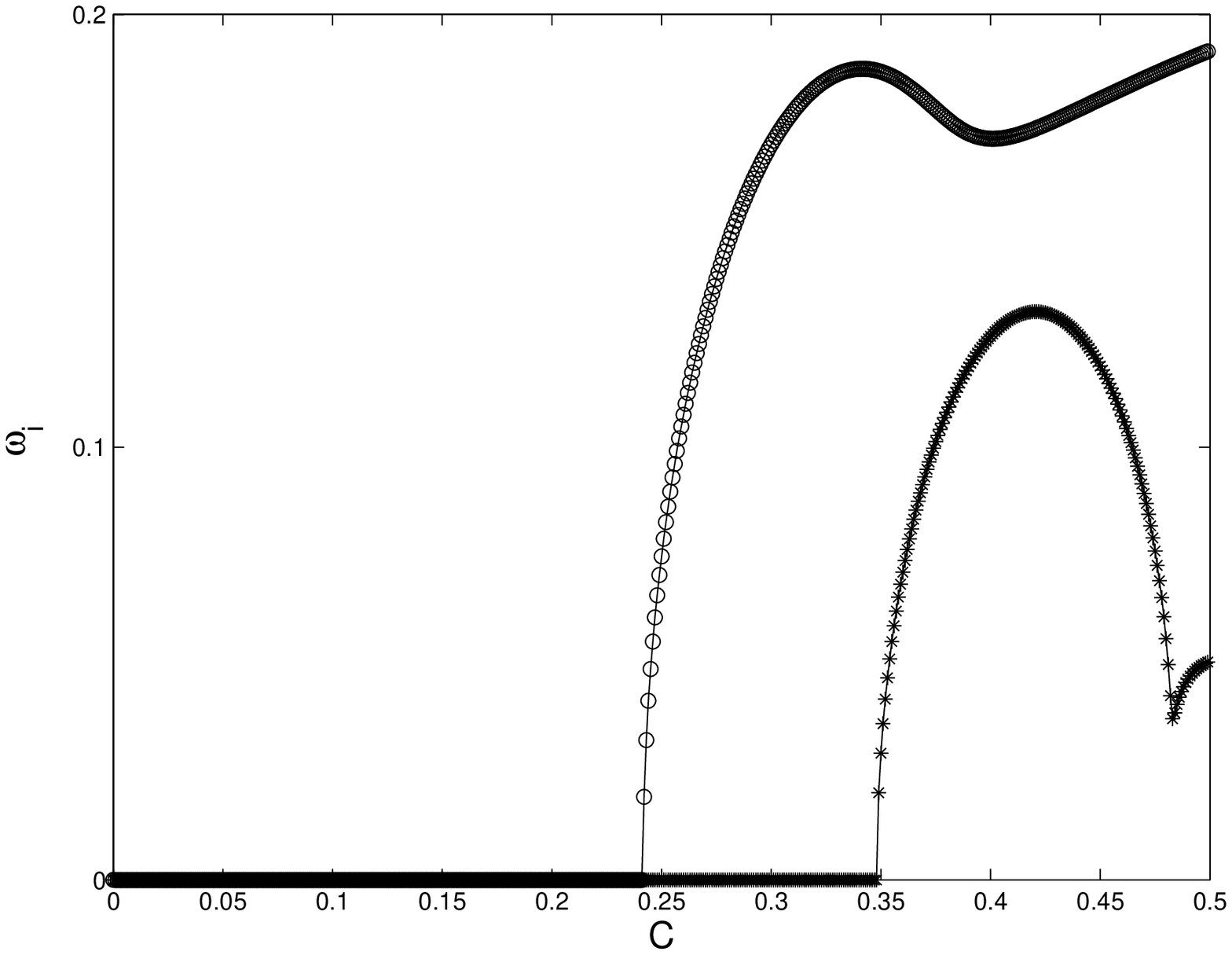}}
\epsfxsize=8.5cm
\centerline{\epsffile{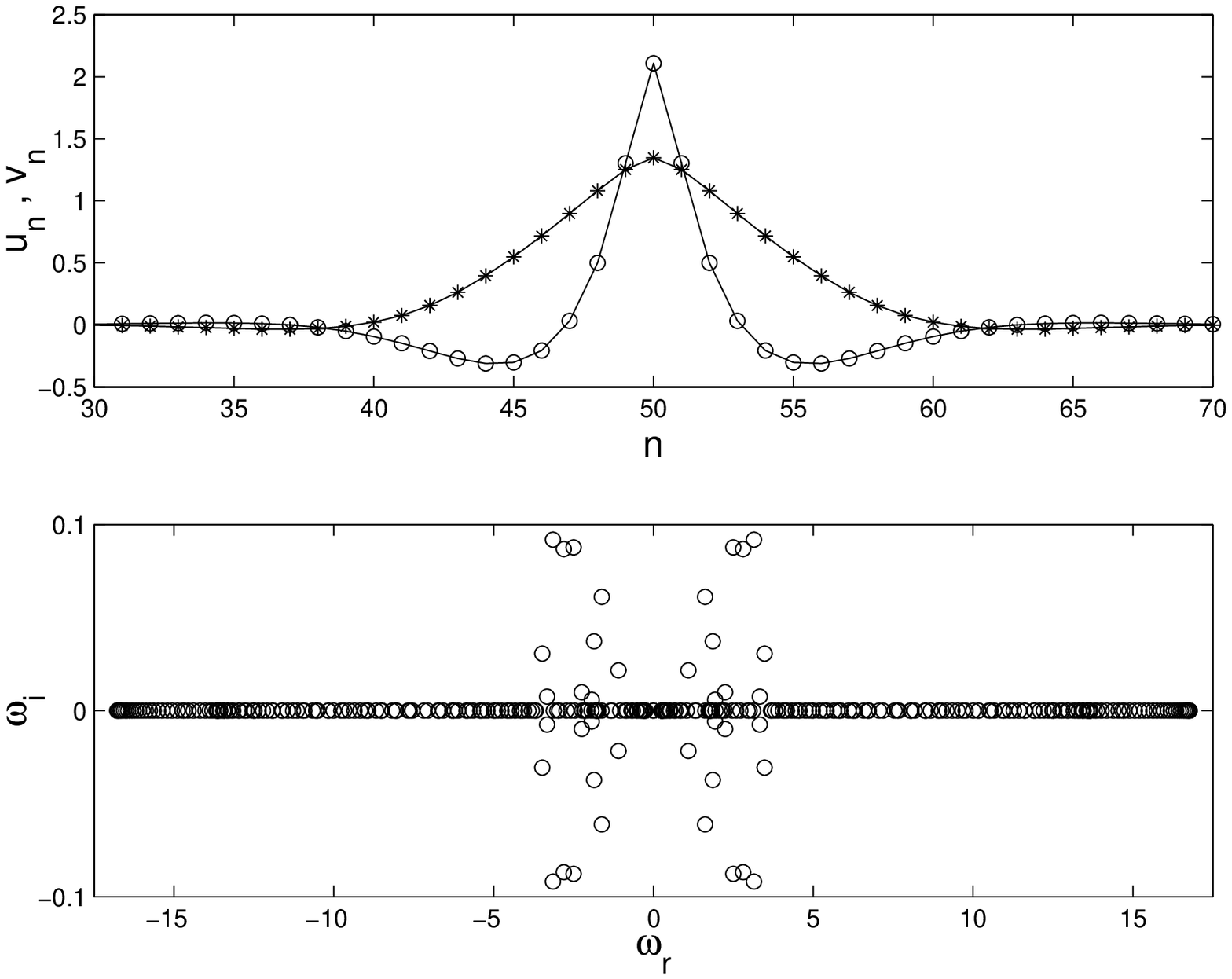}}
\caption{ The branch of the gap-soliton solutions with 
$\Lambda =0.75$, $\protect\kappa =-1$, $\protect\delta =0.9$, and 
$\protect\beta =0$. The
upper part of the figure shows the norms of the two components of the
soliton, and examples of the solutions for $C=0.1$ (stable) and $C=0.4$
(after the onset of the first oscillatory instability). The middle panel
shows the instability growth rates, while the lower part of the figure gives
an example of a solution belonging to this branch, found at a much larger
value of the coupling constant, $C=4$. This solution family extends, as an
unstable one, up to the continuum limit.}
\label{jfig6}
\end{figure}

An example of such a solution for $\Lambda =0.75$, $\kappa =-1$, $\delta =0.9
$, and $\beta =0$ is shown in Fig. \ref{jfig6}. In the AC limit, this branch
starts with the expression (\ref{AC1}). The branch is stable for small $C$,
but then it becomes unstable due to oscillatory instabilities. The first two
instabilities occur at $C=0.242$ and $C=0.349$, as is shown in Fig. \ref
{jfig6}. Past the onset of the instabilities, this branch continues to exist
(as an unstable one) indefinitely with the increase of $C$, and carries over
into an (unstable) GS in the continuum limit. At large values of $C$, the
distinct phonon bands, which are clearly seen in the example of the
eigenvalue spectrum shown for $C=0.4$ in Fig. \ref{jfig6}, eventually
collide and, due to their opposite {\it Krein signs }(see the definition and
discussion of these in Ref. \cite{Krein}), which gives rise to a whole set
of oscillatory instabilities. The result is clearly seen in the example of
the eigenvalue spectrum shown in the bottom panel of Fig. \ref{jfig6} for a
large value of the coupling constant, $C=4$. The characteristic size of the
instability growth rate (largest imaginary part of the eigenvalue) is nearly
the same for $C=0.4$ and $C=4$, in the latter case it being $\approx 0.09$.
Notice, however, that, as the continuum limit is approached, the
instabilities may be suppressed, in a part or completely, by finite-size
effects (for an example of such finite-size restabilization, see Ref. \cite
{JohKiv}).

\begin{figure}[tbp]
\epsfxsize=8.5cm
\centerline{\epsffile{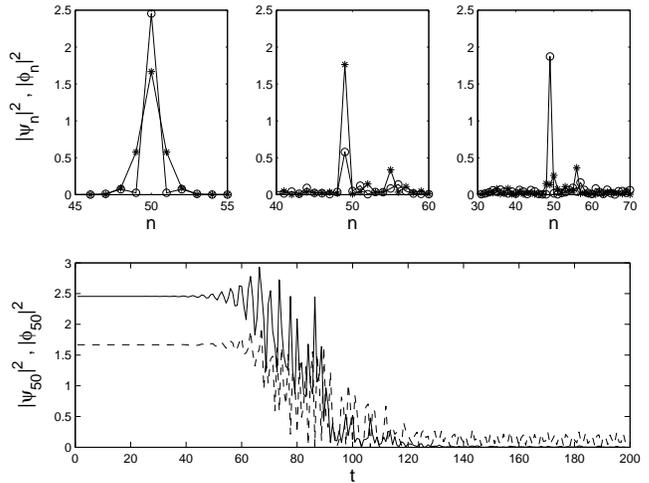}}
\caption{The evolution of the unstable gap soliton belonging to the inner
layer, for $C=0.4$, $\protect\delta=0.9$, $\protect\kappa=-1$, and 
$\protect\beta=0$. The top panels shows the wave field distribution 
at $t=4$ (left), $t=124$ (middle) and $t=132$ (right). 
Symmetry-breaking effects are clearly
visible. A random perturbation of an amplitude $10^{-4}$ was added to the
initial condition in order to catalyze the onset of the instability, which
occurs at $t>40$.}
\label{newfig4}
\end{figure}

The development of the oscillatory instability of GS belonging to the inner
layer is displayed, for $C>0.242$, in Fig. \ref{newfig4} for $C=0.4$, 
$\delta =0.9$, $\kappa =-1$ and $\beta =0$. In this particular case, there
are two oscillatory instabilities whose growth rates are in the interval 
$0.1<\omega _{i}<0.2$. As a result, symmetry breaking occurs, resulting in a
shift of the central position of the soliton (from the site $n=50$ to 
$n=49$). Oscillatory features in the dynamics are also observed in the latter
case, and a small amount of energy is emitted as radiation.

\begin{figure}[tbp]
\epsfxsize=8.5cm
\centerline{\epsffile{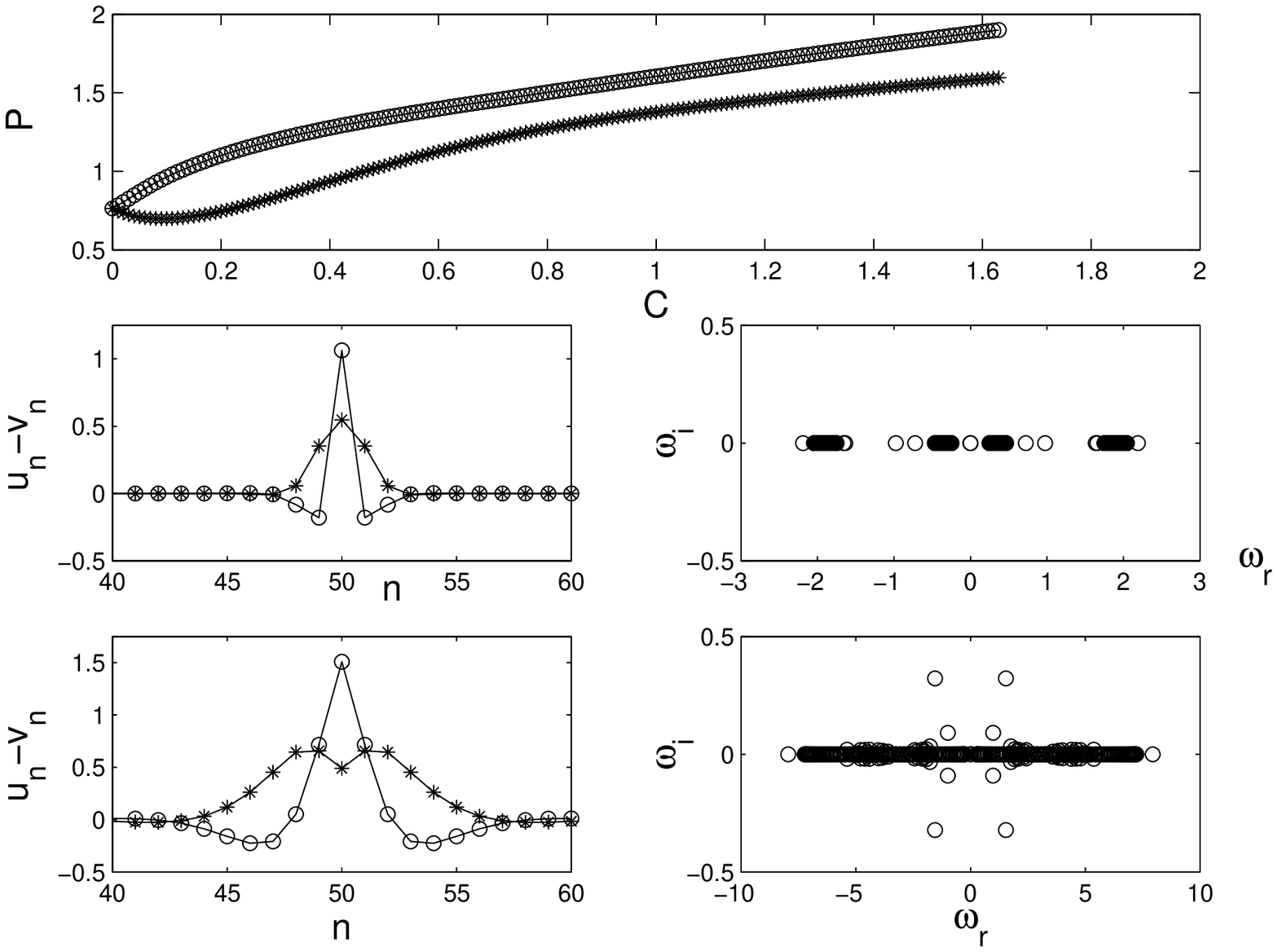}}
\epsfxsize=8.5cm
\centerline{\epsffile{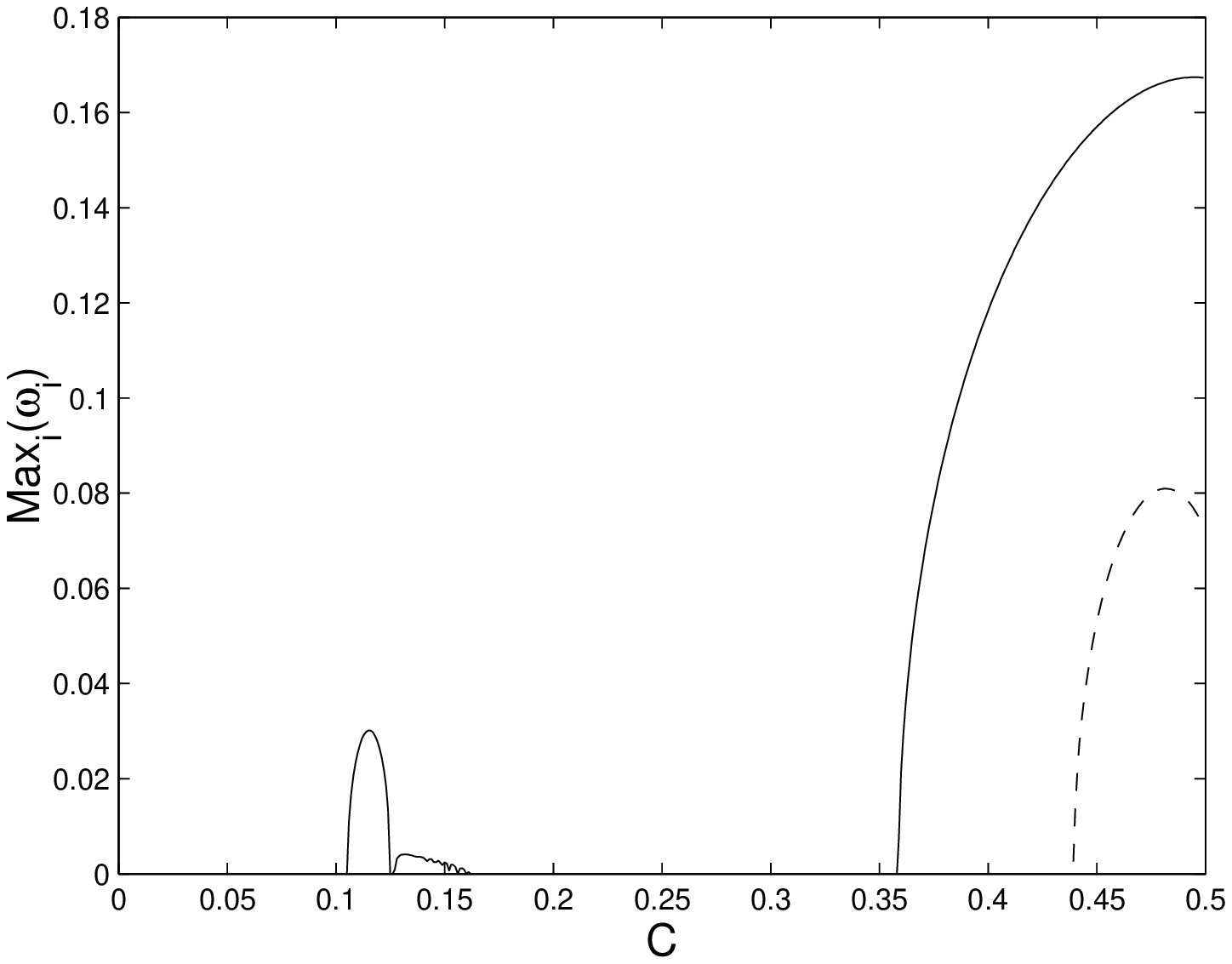}}
\caption{The same as the previous figure, but for $\protect\beta =2$.
Examples of the solutions are shown for $C=0.2$ (upper row, stable) and 
$C=1.6$ (lower row, unstable due to several of oscillatory instabilities).
The bottom panel demonstrates the nonmonotonic evolution of the instability
of this solution with the increase of $C$. }
\label{jfig7}
\end{figure}

Similar results were obtained for smaller values of $\Lambda $, for
instance, $\Lambda =0.25$. It was verified too that this scenario persists
in the presence of the XPM nonlinearity (i.e., for $\beta =2$), as it is
shown in Fig. \ref{jfig7}. In the latter case, the evolution of the
instability with the increase of $C$ is quite interesting, as it is {\em
nonmonotonic}. The instability first arises at $C\approx 0.11$ (due to a
collision between discrete eigenvalues with opposite Krein signs).
Subsequent restabilization takes place at $C\approx 0.16$, but the solutions
are unstable again for $C>0.36,$ and remain unstable thereafter, up to the
continuum limit.

In the case of $\beta =2$, the dynamical development of the oscillatory
instabilities is similar to the $\beta =0$ case, again demonstrating
symmetry-breaking effects.

\subsection{Solutions in the outer layer of the gap}

\begin{figure}[tbp]
\epsfxsize=8.5cm
\centerline{\epsffile{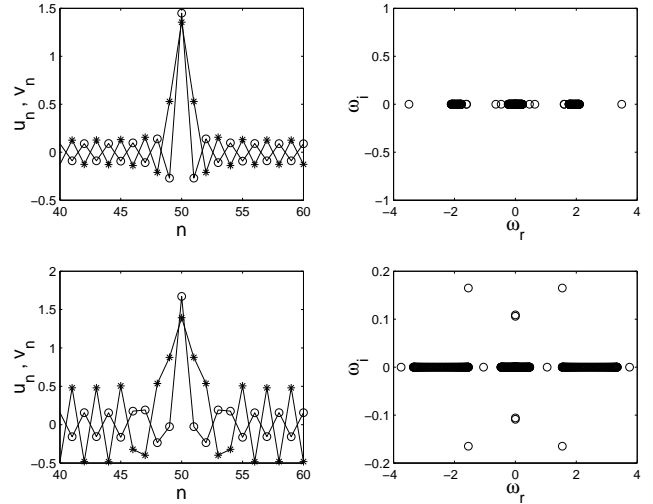}}
\caption{ Solutions with non-decaying oscillatory background for propagation
constants belonging to the outer layer defined by Eq. (\ref{innerouter}).
The top panel shows a stable solution for $C=0.158$, and the bottom panel
shows an unstable one of $C=0.576$. These solutions are unstable for all 
$C>0.341$. }
\label{jfig8}
\end{figure}

In all the cases considered in the previous subsections, the soliton's
propagation constant $\Lambda $ belonged to the inner layer of the gap, see
Eq. (\ref{innerouter}). We have also examined the situation when $\Lambda $
belongs to the outer layer defined in Eq. (\ref{innerouter}) (the outer
layer exists unless $\delta =1$). An example is shown in Fig. \ref{jfig8},
where $\beta =0$, $\kappa =-1$, $\delta =0.1$, and $\Lambda \approx 0.787$
is chosen to be in the middle of the outer layer. In this case, we typically
obtained delocalized solitons, sitting on top of a finite background (they
are sometimes called ``antidark'' solitons). As can be observed from Fig. 
\ref{jfig8}, such solutions may be stable for sufficiently weak coupling,
but become unstable as the continuum limit is approached, although they do
not disappear in this limit (in Ref. \cite{Dave} such delocalized solitons
were found in the continuum counterpart of the present model).

\begin{figure}[tbp]
\epsfxsize=8.5cm
\centerline{\epsffile{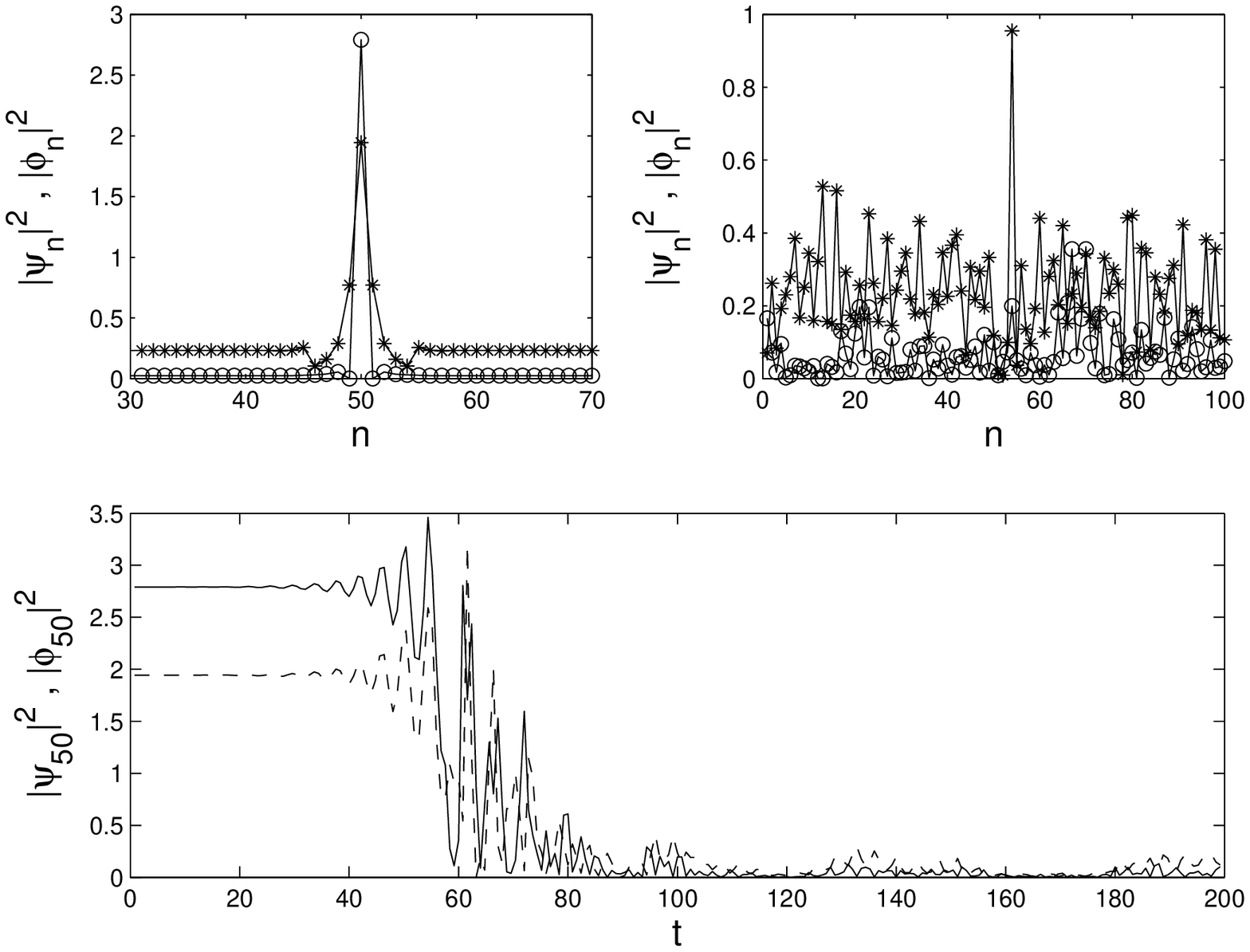}}
\caption{The time evolution of the outer-layer gap solitons for $C=0.549$, 
$\protect\beta=0$, $\protect\kappa=-1$ and $\protect\delta=0.1$. The initial
condition is perturbed by a random uniformly distributed perturbation of an
amplitude $10^{-4}$. The result of the instability is the excitation of
background oscillations, as well as a shift of the soliton's peak from its
original position. The top left and right spatial profiles correspond to 
$t=4 $ and $t=200$, respectively. }
\label{newfig5}
\end{figure}

The instability development in the case of the outer-layer GSs is
demonstrated, for $C=0.549$, $\beta =0$, $\kappa =-1$ and $\delta =0.1$, in
Fig. \ref{newfig5}. In this case, the non-vanishing background is also
perturbed by the instability, resulting in, plausibly, chaotic oscillations
throughout the lattice. Symmetry-breaking effects, which shift the central
peak from its original position, are observed too in this case.

\subsection{Stabilization of the gap solitons by mismatch}

The above considerations show that, inside the inner layer of the gap, it is
easy to identify families of soliton solutions that persist in the continuum
limit as $C\rightarrow \infty $. However, all the examples considered above
showed that the solutions get destabilized at finite $C$ and remain unstable
with the subsequent increase of $C$. Therefore, a challenging problem is to
find solution families that would remain stable for large values of $C$.

\begin{figure}[tbp]
\epsfxsize=8.5cm
\centerline{\epsffile{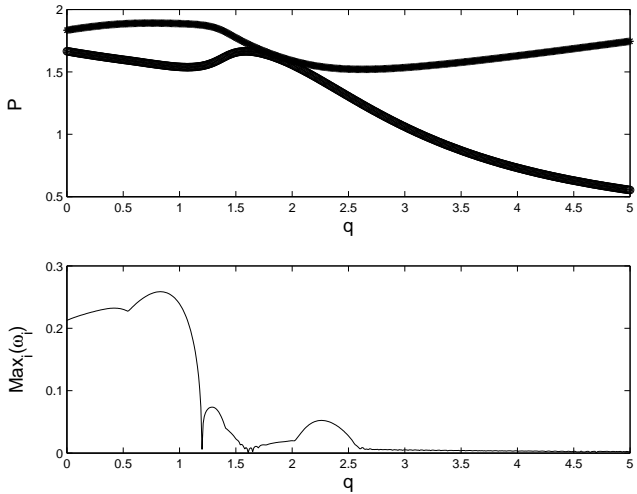}}
\epsfxsize=8.5cm
\centerline{\epsffile{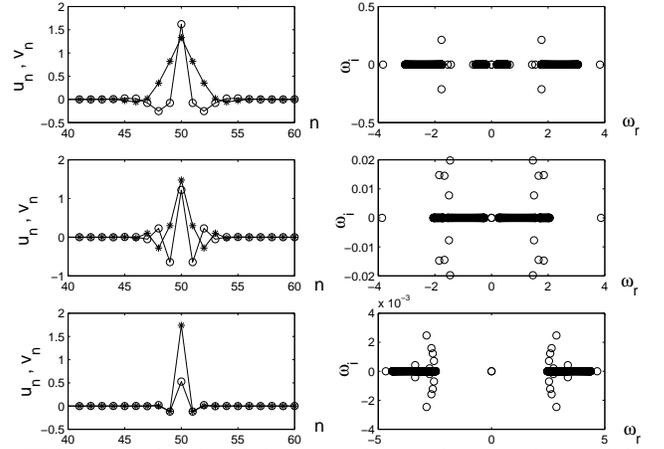}}
\caption{A family of the gap-soliton solutions obtained for fixed values 
$C=0.5$, $\protect\kappa =-1$, $\protect\delta =0.5$, $\Lambda =0.75$, and 
$\protect\beta =0$, by continuation to positive values of the mismatch
parameter $q$. The top panel and the one beneath it show the evolution of
the norms of the two components of the solution, and of the largest
instability growth rate, with the increase of $q$. Other panels show
examples of the solution (as usual, in terms of profiles of the two
components and linear stability eigenvalues) for $q=0$, $q=2$, and $q=5$,
from top to bottom.}
\label{mismatch+}
\end{figure}

\begin{figure}[tbp]
\epsfxsize=8.5cm
\centerline{\epsffile{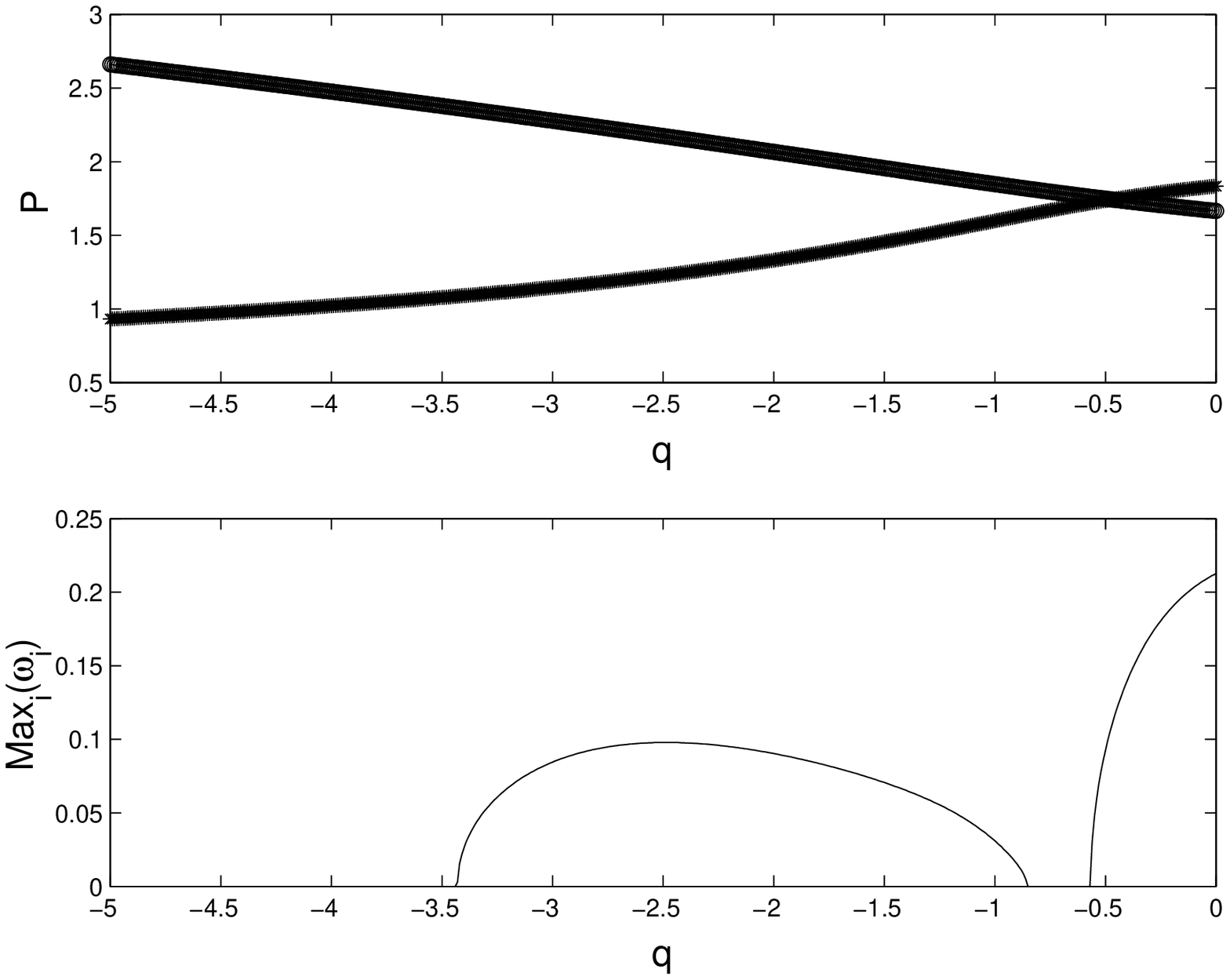}}
\epsfxsize=8.5cm
\centerline{\epsffile{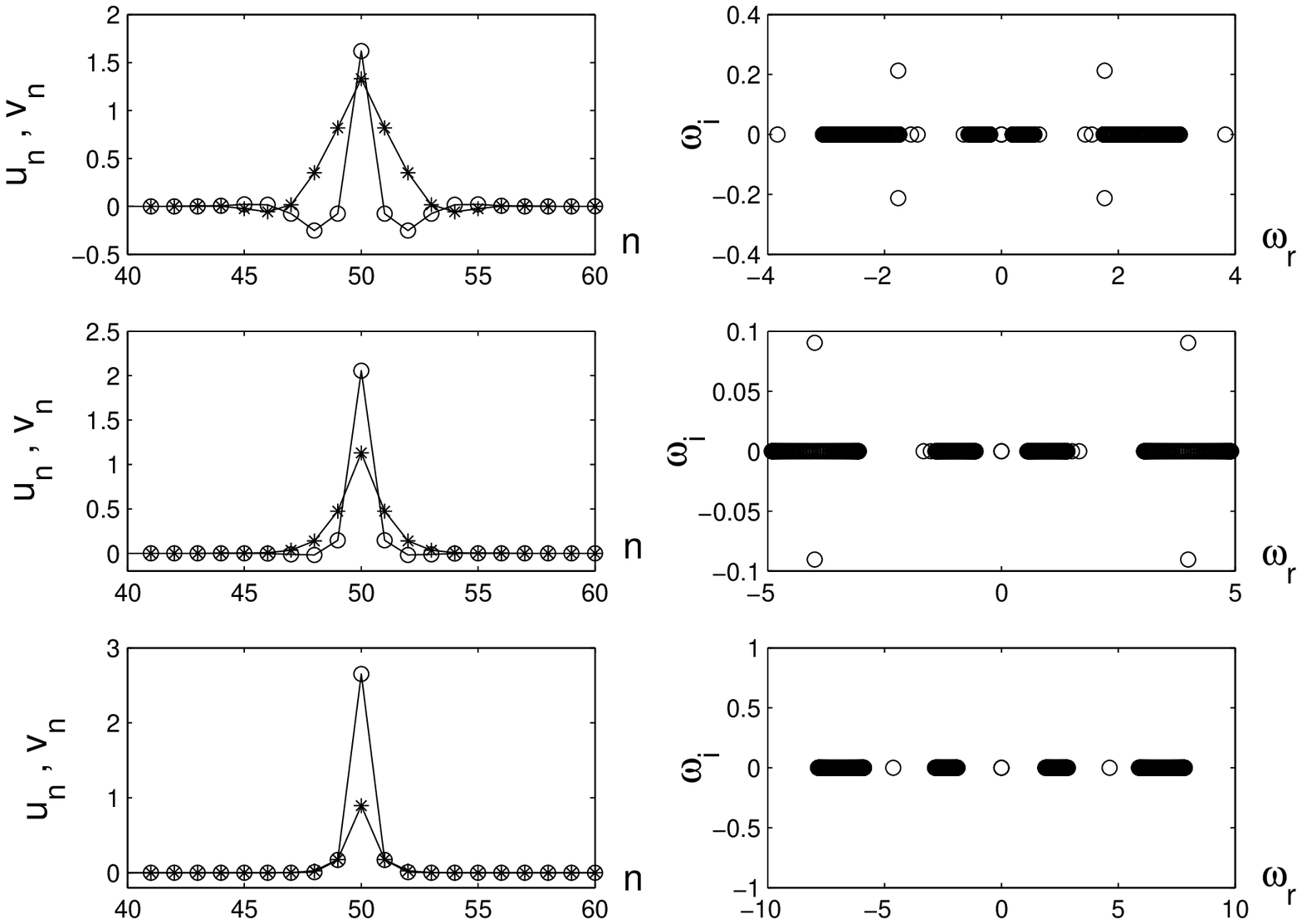}}
\caption{The same as in Fig. \ref{mismatch+}, but for negative values of the
mismatch. Examples of the solutions are given for $q=0$, $q=-2$, and $q=-5$,
from top to bottom. A random perturbation of an amplitude $10^{-4}$ was used
in this case.}
\label{mismatch-}
\end{figure}

In fact, the introduction of a finite mismatch $q$ (recall it was set equal
to zero in all the examples considered above) may easily stabilize the
discrete GSs. To this end, we pick up a typical example, with $C=0.5$, 
$\kappa =-1$, $\delta =0.5$, $\Lambda =0.75$, and $\beta =0$, when the GS
exists but is definitely unstable in the absence of the mismatch. Figures 
\ref{mismatch+} and \ref{mismatch-} show the effect of positive and negative
values of the mismatch on the solitons. As is seen, large values of the
positive mismatch can make the instability very weak, but cannot completely
eliminate it. However, sufficiently large negative mismatch readily makes
the solitons {\em truly stable}. Thus, adding the negative mismatch is the
simplest way to stabilize the solitons at large $C$, which is not
surprising, as Eq. (\ref{gap}) demonstrates that the negative mismatch makes
the gap broader.

\begin{figure}[tbp]
\epsfxsize=8.5cm
\centerline{\epsffile{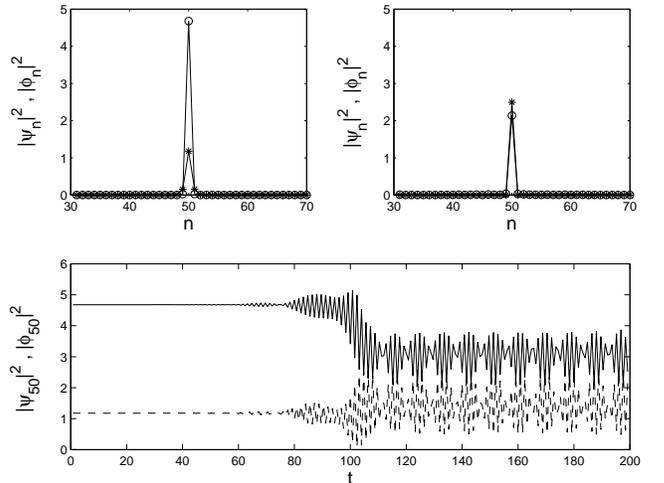}}
\caption{The dynamical evolution in the unstable case with $C=0.5$, 
$\protect\kappa=-1$, $\protect\beta=0$, $\protect\delta=0.5$ 
and $q=1$ (positive
mismatch). The top left and right panels correspond to $t=4$ and $t=396$,
respectively. A complex pattern of the amplitude evolution is observed in
this case. The initial condition contains a random perturbation with an
amplitude $5 \times 10^{-5}$.}
\label{newfig6}
\end{figure}

As an example of the dynamical evolution of unstable solitons in the case of
positive mismatch, in Fig. \ref{newfig6} we display the case of $C=0.5$, 
$\kappa =-1$, $\beta =0$, $\delta =0.5$ and $q=1$. In this case, the
evolution leads to the establishment of a breather with a rather complex
dynamical behavior.

\begin{figure}[tbp]
\epsfxsize=8.5cm
\centerline{\epsffile{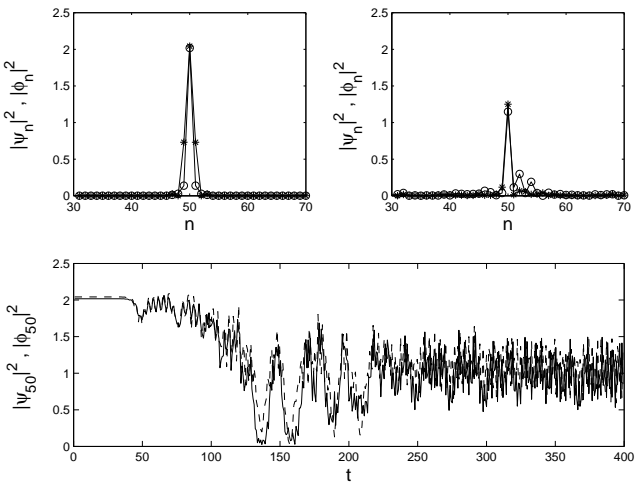}}
\caption{The evolution of the unstable discrete gap soliton in with $C=0.5$, 
$\protect\kappa =-1$, $\protect\beta =0$, $\protect\delta =0.5$ and $q=-2.5$
(negative mismatch). The top left and right panels show the field profiles
for $t=4$ and $t=196$, respectively. The time evolution of the
amplitudes is shown in the bottom panel.}
\label{newfig7}
\end{figure}

The instability development in the case of negative mismatch, $q=-2.5$, is
demonstrated in Fig. \ref{newfig7}. A localized breather with quasi-periodic
intrinsic dynamics is observed in this case as an eventual state.

\subsection{Discrete counterparts of gap solitons from the Bragg-grating
model}

As  was shown in the introduction, the particular case of Eqs. 
(\ref{zeq1}) and (\ref{zeq2}) with $\kappa =1$, $\delta =1$, 
and $\beta =2$ may be
interpreted, with regard to the transformation (\ref{stagger}), as a
discretization of the standard Bragg-grating system (\ref{BG}). This
continuum model gives rise to a family of exact GS solutions \cite{Exact}, 
\begin{equation}
\Psi =U(x)\,\exp \left( -it\cos \theta \right) ,\Phi =V(x)\,\exp \left(
-it\cos \theta \right) ,  \label{stationary}
\end{equation}
\begin{equation}
U(x)=\frac{\sin {\rm \,}\theta }{\sqrt{3}}\,{\rm sech}\left( x\sin {\rm \,}
\theta -\frac{i}{2}\theta \right) ,\,V=-U^{\ast },  \label{exact}
\end{equation}
where the real parameter $\theta $ takes values $0<\theta <\pi $. A part of
this interval, $0<\theta <\theta _{{\rm cr}}\approx 1.01\left( \pi /2\right) 
$, is filled with stable solitons \cite{stableBG}, while the remaining part
contains unstable ones.

All the discrete GSs considered above are {\em not} counterparts of the
continuum solitons given by Eqs. (\ref{stationary}) and (\ref{exact}).
Establishing a direct correspondence between the latter ones and discrete
solitons of Eqs. (\ref{zeq1}) and (\ref{zeq2}) is complicated by two
problems: the transformation (\ref{stagger}) does not have a continuum
limit, and {\em real} symmetric or anti-symmetric GSs with $\left| \Lambda
\right| <\kappa $ do not exist in the AC\ limit, as is seen from Eqs. (\ref
{AC1}) and (\ref{AC2}), i.e., the usual starting point of the analysis is
not available in this case.

\begin{figure}[tbp]
\epsfxsize=8.5cm
\centerline{\epsffile{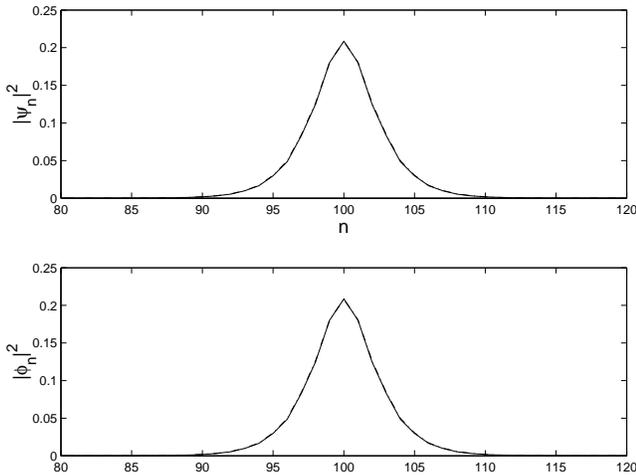}}
\caption{Absolute values of the fields in a new {\it complex} discrete
gap-soliton solution, which was obtained, at $C=1.24$, for the
formal discretization of the Bragg-grating gap-soliton model. 
Then, the transformation (\ref{stagger}) was applied to this solution and
it was used as an initial condition of the Newton for finding a solution
of Eqs. (\ref{zeq1})-(\ref{zeq2}). The continuous and dashed
lines (which completely oevrlap) show the resulting profiles generated by the
above-mentioned procedure, one which is an exact solution for the 
discretization of the Bragg-grating model and one which is an exact
solution (identical in norm due to the nature of (\ref{stagger}))
of Eqs. (\ref{zeq1})-(\ref{zeq2}).}
\label{newsoliton}
\end{figure}

\begin{figure}[tbp]
\epsfxsize=8.5cm
\centerline{\epsffile{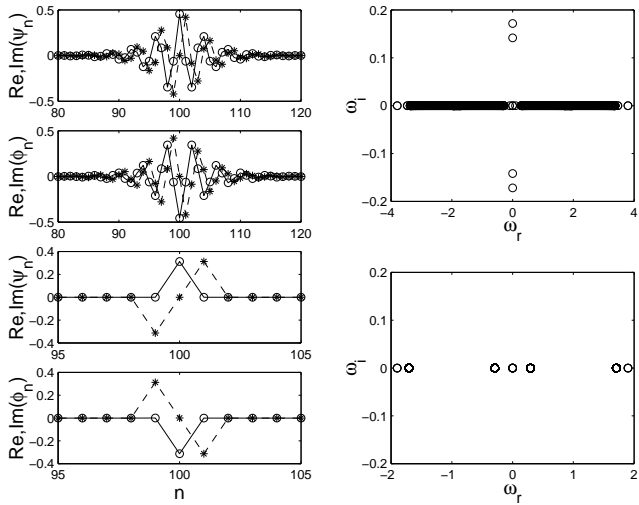}}
\caption{Left panels show profiles of real and imaginary parts of the fields 
$\psi_n$ and $\phi_n$ in the discrete counterpart of the 
Bragg-grating gap soliton
from Fig. \ref{newsoliton}. Right panels show stability eigenfrequencies 
for the
same soliton. The upper and lower parts of the figure pertain to the soliton
at $C=1.24$ (the same value as in Fig. \ref{newsoliton}), and to its
continuation to the anti-continuum limit, $C=0$. In the left panels,
the circles (joined by solid lines) refer to the real parts, while the 
stars (connected by dashed lines) correspond to the imaginary parts of the
corresponding shown fields.}
\label{newsolitonAC}
\end{figure}

We have considered the discrete analogs of the Bragg-grating
GSs in the following way. First, we took a formal discrete counterpart of
the waveforms (\ref{stationary}) and (\ref{exact}), and used them
to obtain exact solutions of the (formal) discretization of the 
Bragg-grating model of Eq. (\ref{BG}). Then the
transformation (\ref{stagger}) was applied to these solutions, 
and the thus obtained expressions
were used as an initial guess for finding a numerically exact stationary
solution of Eqs. (\ref{zeq1})-(\ref{zeq2}). 
This procedure naturally generates new solitons, a crucial
difference of which from all the types considered above is that they are 
{\em truly complex} solutions, see examples in Figs. \ref{newsoliton} and 
\ref{newsolitonAC}. In this case, we have used $\theta=\pi/4$.

Then, the solution was numerically continued, decreasing $C$, {\em back} to
the AC limit, in order to identify its AC ``stem''. The result is shown in
Fig. \ref{newsolitonAC}. Obviously, this AC state is very different from
all those considered above (in particular, it is complex).

\begin{figure}[tbp]
\epsfxsize=8.5cm
\centerline{\epsffile{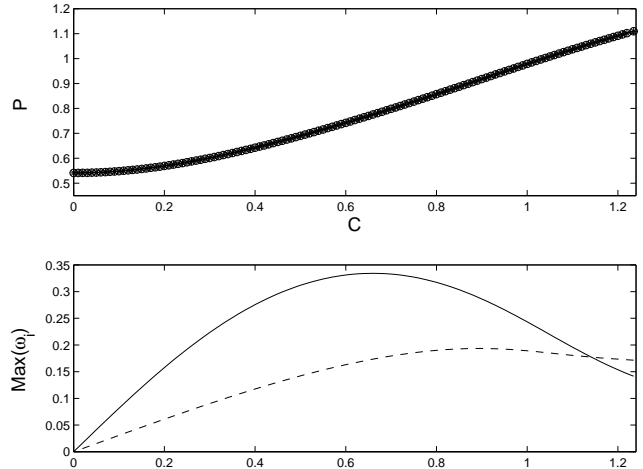}}
\caption{The norm of the discrete counterpart of the Bragg-grating gap
soliton of the model of Eqs. (\ref{zeq1})-(\ref{zeq2}) (upper panel), 
and its two unstable (imaginary) eigenfrequencies (lower panel shown
by solid and dashed lines respectively), vs.
the coupling constant. The continuation of the figure to larger values of $C$
shows that the soliton becomes asymptotically stable as $C\to\infty$.}
\label{newsolitonStability}
\end{figure}

Finally, linear stability eigenvalues 
were calculated for this new branch of the
discrete GSs. The result (see Fig. \ref{newsolitonStability}) is that this
branch is unstable for all finite values of $C$, getting asymptotically
stable in both limits $C\rightarrow 0$ and $C\rightarrow \infty $ (large
values of $C$ are not shown in Fig. \ref{newsolitonStability}); the 
stability regained in the latter limit complies with the above-mentioned
finding that a subfamily of the continuum Bragg-grating gap solitons
are dynamically stable. Notice that
the natural norm of the continuum soliton differs from that of the discrete
one, given by Eq. (\ref{P}), by an additional multiplier $C^{-1/2}$ (which
is proportional to the effective lattice spacing). We have checked that the
thus renormalized norm of the soliton converges as $C\rightarrow \infty $,
although data for large $C$ is not displayed in Fig. 
\ref{newsolitonStability}.

\section{Solitons in the model with the quadratic nonlinearity}

Stationary solutions of the SHG system (\ref{jeq12}) and (\ref{jeq13}) are
looked for in an obvious form, cf. Eqs.\ (\ref{zeq4}): 
\begin{equation}
\psi _{n}=e^{i\Lambda t}u_{n},\,\phi _{n}=e^{2i\Lambda t}v_{n},
\label{jeq15a}
\end{equation}
and in this case we only consider the (most characteristic) 
case $\delta =1$. The linearization of Eqs. (\ref{jeq12}) 
and (\ref{jeq13}) demonstrates
that one may expect termination of a soliton-solution branch, due to its
collision with the phonon band, at (or close to) the point 
\begin{equation}
\Lambda =\kappa /4+C,  \label{jeq15}
\end{equation}
and the gap between two phonon bands is 
\begin{equation}
0<\Lambda <\kappa /4  \label{chi2gap}
\end{equation}
(it exists only if $\kappa >0$).

Stationary solutions were constructed, again, by means of continuation
starting from the AC limit, where the excitation localized on a single site
of the lattice assumes the form 
\begin{eqnarray}
v_{n_{0}} &=&\Lambda ,  \label{jeq16} \\
u_{n_{0}} &=&\pm \sqrt{v_{n_{0}}(4\Lambda -\kappa )}.  \label{jeq17}
\end{eqnarray}
Note that solutions with the propagation constant belonging to the gap (\ref
{chi2gap}) do not exist close to the AC limit. Indeed, the AC expression 
(\ref{jeq17}) shows that a necessary condition for its existence is $4\Lambda
>\kappa $. On the other hand, $\Lambda $ stays in the gap (\ref{chi2gap}) if 
$4\Lambda <\kappa $, so the two conditions are incompatible.

\begin{figure}[tbp]
\epsfxsize=8.5cm
\centerline{\epsffile{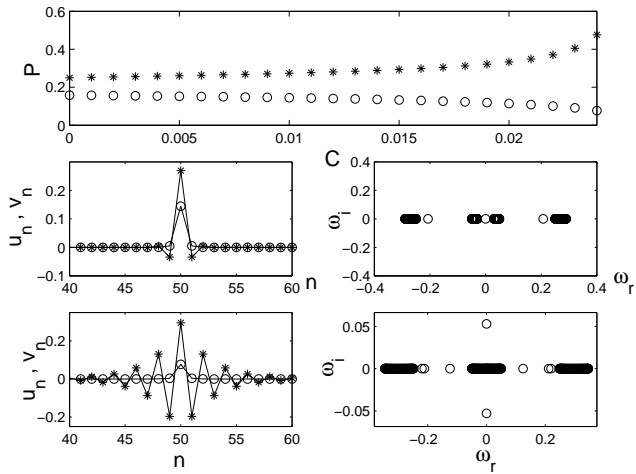}}
\epsfxsize=8.5cm
\centerline{\epsffile{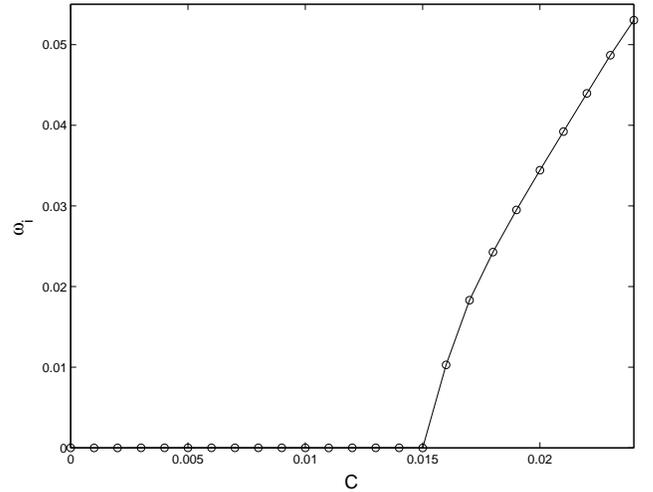}}
\caption{The family of the discrete SHG solitons found for $\protect\kappa
=0.9$, $\protect\delta =1$, and $\Lambda =0.25$. The top panel shows the
evolution of the norms of the fundamental- (circles) and second-frequency
(stars) components of the soliton, which are defined the same way as in Eq. 
(\ref{P}), with the increase of $C$. Examples of solutions are displayed for 
$C=0.01$, below the instability threshold, which is $C_{{\rm cr}}=0.015$ (the
upper row), and for $C=0.024$, just prior to the termination of the branch
at $C=0.025$ (the lower row). The bottom panel shows the evolution of the
instability growth rate (imaginary part of the most unstable eigenvalue).}
\label{jfig9}
\end{figure}

A typical example of a numerically found soliton branch is shown, for the
value of the mismatch parameter $\kappa =0.9$, in Fig. \ref{jfig9}. This
solution family is fixed by choosing $\Lambda =0.25$, and starting from the
expressions (\ref{jeq16}) and (\ref{jeq17}) with the upper sign. It is seen
that the branch is stable for $C<0.015$, but then it becomes unstable, and
eventually terminates at $C\approx 0.025$, in very good agreement with the
prediction of Eq. (\ref{jeq15}).

\begin{figure}[tbp]
\epsfxsize=8.5cm
\centerline{\epsffile{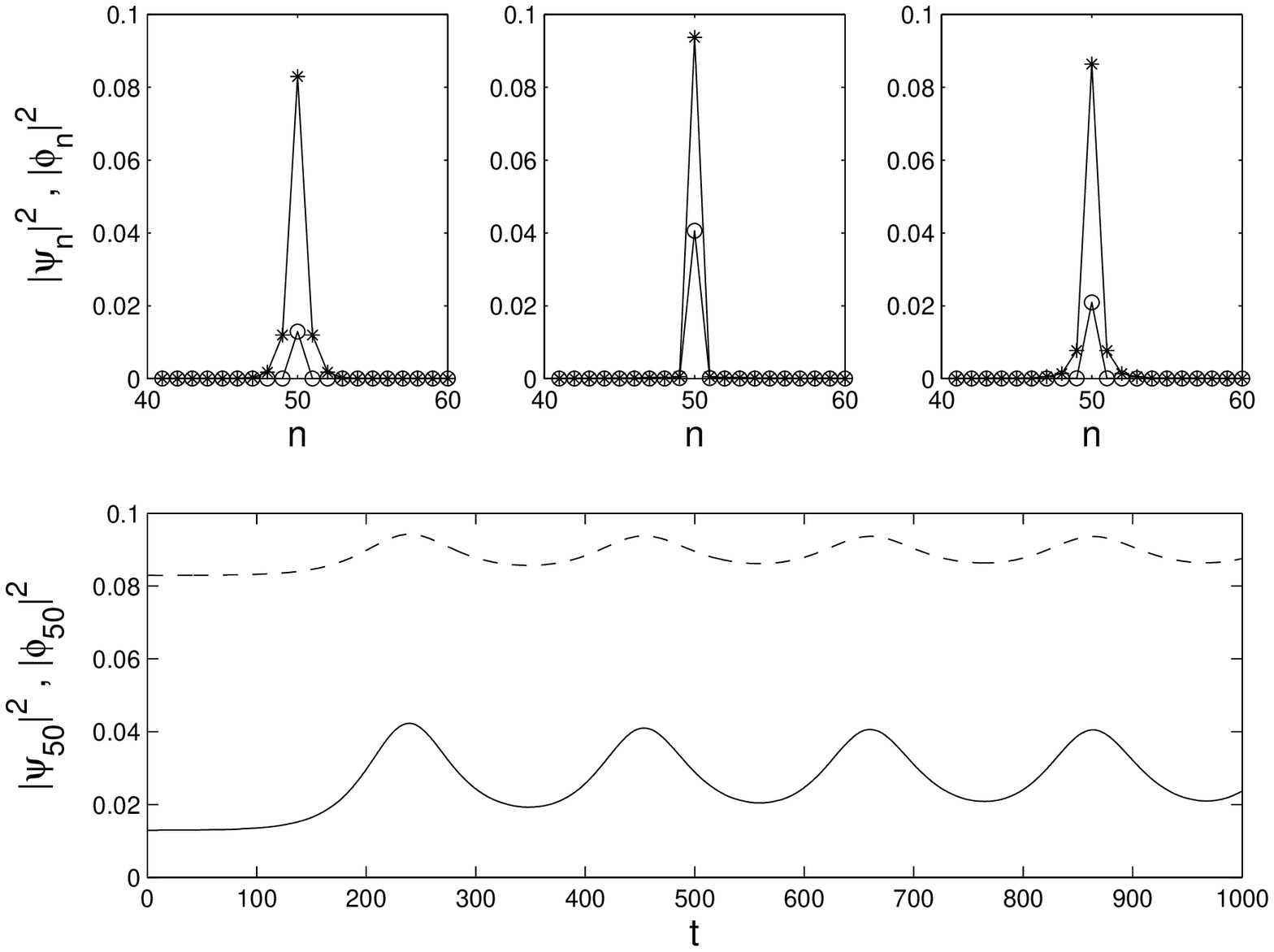}}
\caption{Simulations of the instability development for the discrete SHG
soliton in the case with $C=0.02$, $\protect\kappa=0.9$ and $\protect\delta=1
$. The circles in the spatial profiles of the top panels denote the
fundamental, and the stars denote the second harmonic. The left panel
corresponds to $t=4$, the middle to $t=660$ (close to the point where the
oscillating amplitude attains its maximum), and the right to $760$ (close to
a minimum-amplitude point). The bottom panel shows the oscillatory behavior
at the central site for the fundamental (solid line) and second harmonics
(dashed line).}
\label{newfig8}
\end{figure}

Development of the instability of this SHG soliton branch for $C>0.015$ was
numerically examined through direct simulations, results of which are
presented in Fig. \ref{newfig8}, for the case of $C=0.02$, the corresponding
instability growth rate being $\approx 0.03$. A random uniformly distributed
perturbation of an amplitude $10^{-4}$ was added to the initial condition to
accelerate the onset of the instability. The eventual result of the
instability is the appearance of a breather with very regular periodic
intrinsic vibrations.

\begin{figure}[tbp]
\epsfxsize=8.5cm
\centerline{\epsffile{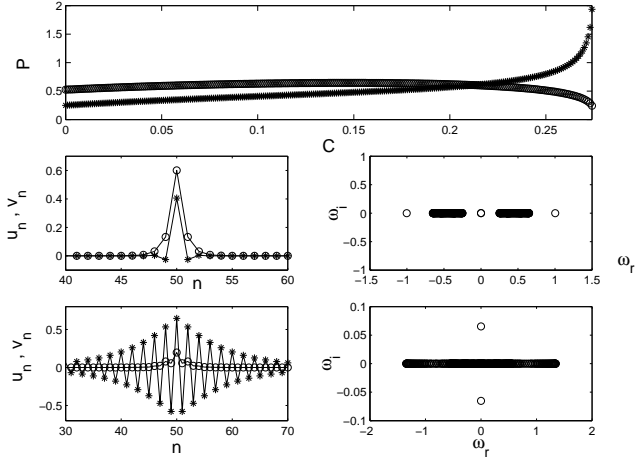}}
\caption{ The same as in the upper part of Fig. \ref{newfig8}, but for 
$\protect\kappa=-0.1$. The middle panel shows an example of the solution for 
$C=0.1$, and the bottom panel shows an example for $C=0.274$, just prior to
the termination of the branch (which happens at $C=0.275$).}
\label{jfig10}
\end{figure}

A similar result for the case without a gap in the phonon spectrum [see Eq. 
(\ref{chi2gap})] is shown in Fig. \ref{jfig10} for $\kappa =-0.1$, the other
parameters being the same as in the previous case. This time, the branch
terminates at $C\approx 0.275$, once again in complete agreement with the
prediction of Eq. (\ref{jeq15}).

\begin{figure}[tbp]
\epsfxsize=8.5cm
\centerline{\epsffile{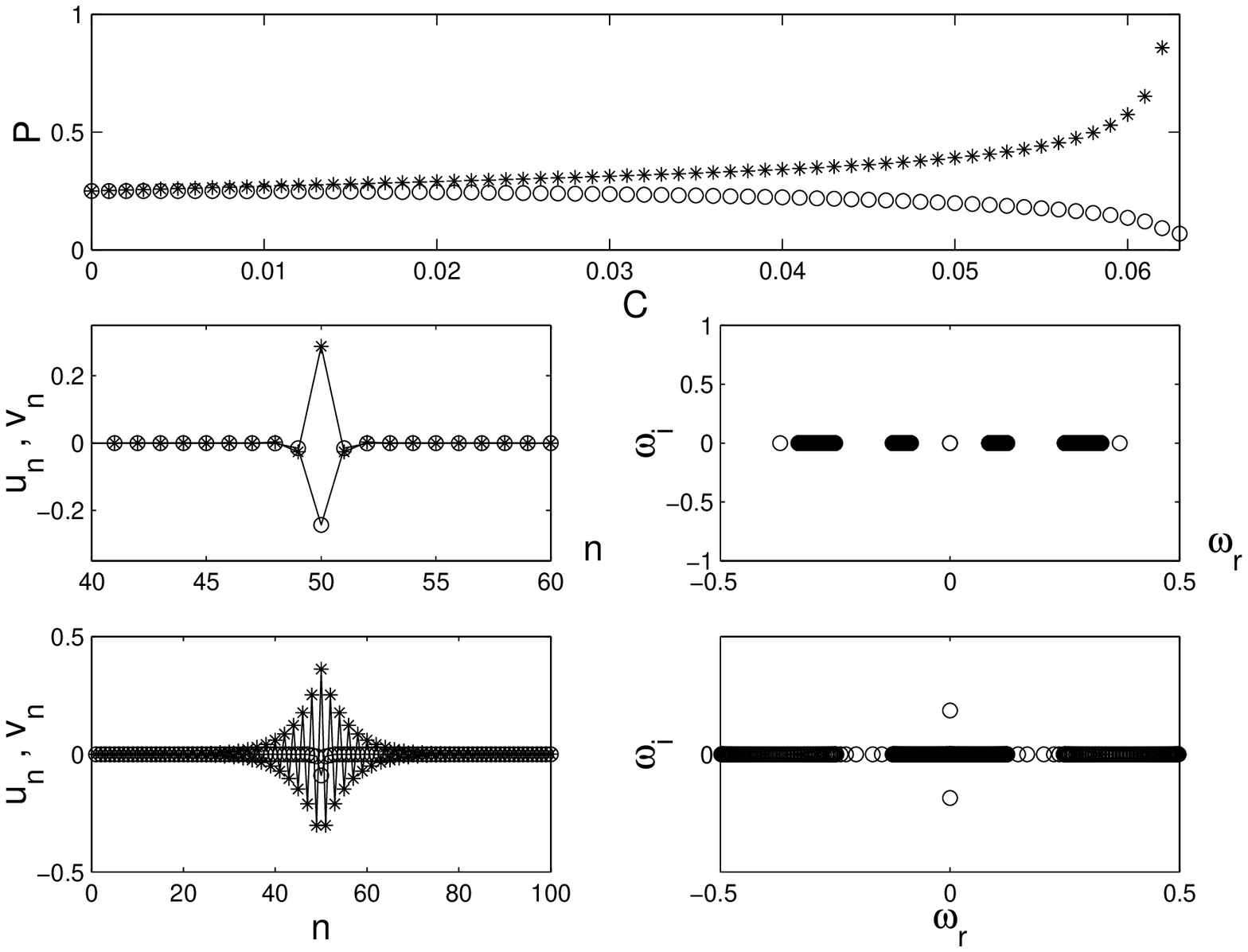}}
\epsfxsize=8.5cm
\centerline{\epsffile{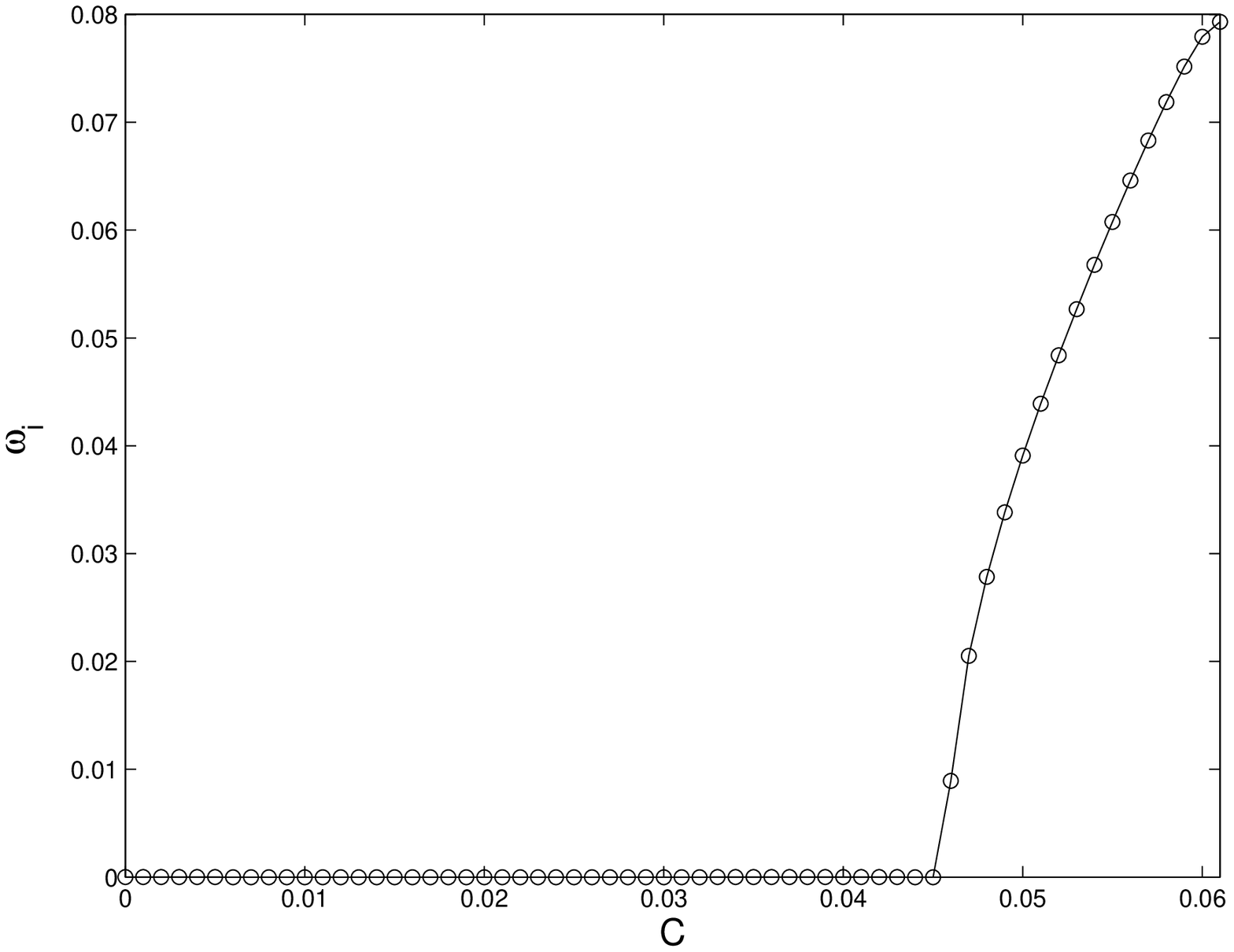}}
\caption{ The same as in Fig. \ref{jfig9}, but for the lower sign in Eq. 
(\ref{jeq17}). The middle panel of the top subplot displays an example of a
stable solution for $C=0.01$, while the bottom panel shows a solution for 
$C=0.061$, close to the termination of the branch. In this case, 
$\protect\kappa =0.75$, $\protect\delta =1$, and $\Lambda =0.25$.}
\label{jfig11}
\end{figure}

Lastly, another characteristic branch of solutions can be constructed
starting from the pattern given by Eqs. (\ref{jeq17}) with the lower sign.
This solution family is displayed in Fig. \ref{jfig11}, for $\kappa =0.75$, 
$\delta =1$, and $\Lambda =0.25$. The branch is stable for sufficiently weak
coupling, but then it becomes unstable for $C>0.046$. The branch disappears
colliding with the phonon band at $C\approx 0.062$, once again in full
agreement with the prediction of Eq. (\ref{jeq15}).

\begin{figure}[tbp]
\epsfxsize=8.5cm
\centerline{\epsffile{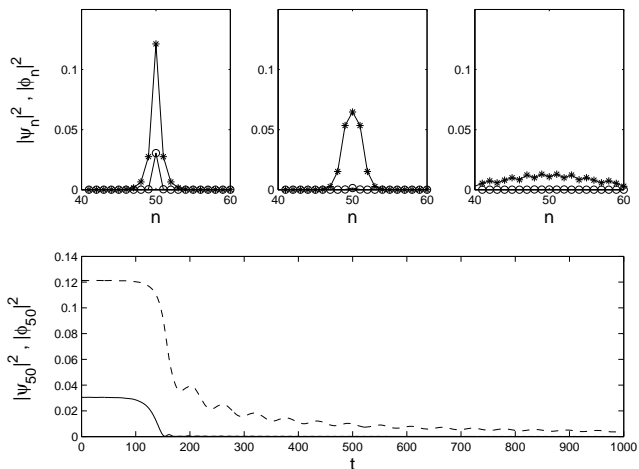}}
\caption{The evolution of the unstable SHG soliton in the case of $C=0.055$, 
$\protect\kappa=0.75$ and $\protect\delta=1$. The top left, middle, and
right panels correspond to the profiles at $t=4$, $t=160$, and $t=400$. The
bottom panel, as before, shows the time evolution of amplitudes at the
central site.}
\label{newfig9}
\end{figure}

An example of the development of instability of the present solution, that
takes place at $C>0.046$, is shown in Fig. \ref{newfig9} for $C=0.055$ 
($\kappa =0.75$; $\delta =1$). A random initial perturbation with an amplitude 
$10^{-4}$ was added to the initial condition in this case. As is seen, the
evolution results in complete destruction of the pulse into small-amplitude
radiation waves.

\section{Conclusion}

In this work, we have introduced a model which includes two nonlinear
dynamical chains with linear and nonlinear couplings between them, and
opposite signs of the discrete diffraction inside the chains. In the case of
the cubic nonlinearity, the model finds two distinct interpretations in
terms of nonlinear optical waveguide arrays, based on the
diffraction-management concept. A continuum limit of the model is tantamount
to a dual-core nonlinear optical fiber with opposite signs of dispersion in
the two cores. Simultaneously, the system is equivalent to a formal
discretization of the standard model of Bragg-grating solitons. A
straightforward discrete second-harmonic-generation [$\chi ^{(2)}$] model,
with opposite signs of the diffractions at the fundamental and second
harmonics, was introduced too. Starting from the anti-continuum (AC) limit
and gradually increasing the coupling constant, soliton solutions in the 
$\chi ^{(3)}$ model were found, both above the phonon band and inside the
gap. Above the gap, the solitons may be stable as long as they exist, but
with transition to the continuum limit they inevitably disappear. On the
contrary, solitons in the gap persist all the way up to the continuum limit.
In the zero-mismatch case, they always become unstable before reaching the
continuum limit, but finite mismatch may strongly stabilize them. A separate
procedure had to be developed to search for discrete counterparts of the
well-known Bragg-grating gap solitons. As a result, it was found that
discrete solitons of this type exist at all values of the coupling constant 
$C$, but they appear to be  
stable solely in the limit cases $C=0$ and $C=\infty $.
Solitons were also found in the $\chi ^{(2)}$ model. They too start as
stable solutions, but then lose their stability.

In the cases when the solitons were found to be unstable, simulations of
their dynamical evolution reveal a variety of different scenarios. These
include establishment of localized breathers featuring periodic,
quasi-periodic, or very complex intrinsic dynamics, or destruction of one
component of the soliton, as well as symmetry-breaking effects, and even
complete decay of both components into small-amplitude radiation. The
outcome depends on the type of the nonlinearity (cubic or quadratic), and on
the nature of the unstable solution.

\section*{Acknowledgements}

B.A.M. acknowledges hospitality of the Department of Applied Mathematics at
the University of Colorado, Boulder, and of the Center for Nonlinear Studies
at the Los Alamos National Laboratory. P.G.K. gratefully acknowledges the
hospitality of the Center for Nonlinear Studies of the Los Alamos National
Laboratory, as well as partial support from the University of Massachusetts
through a Faculty Research Grant, from the Clay Mathematics Institute
through a Special Project Prize Fellowship and from NSF-DMS-0204585. 
Work at Los Alamos is supported
by the U.S. Department of Energy, under contract W-7405-ENG-36.

We appreciate valuable discussions with M.J. Ablowitz and A. Aceves.

\end{multicols}

\end{document}